\documentstyle[12pt,epsf]{article}

\newcommand{\del}{\partial}
\newcommand{\Gam}{\Gamma}
\newcommand{\dis}{\displaystyle}
\newcommand{\ba}{\begin{eqnarray}}
\newcommand{\ea}{\end{eqnarray}}
\newcommand{\non}{\nonumber\\}
\newcommand{\nom}{\nonumber}
\newcommand{\vpi}{\varpi}
\newcommand{\tpi}{\tilde{\varpi}}
\newcommand{\kae}{\mbox{K\"ahler}}

\newcommand{\wa}{\sum_{\stackrel{\scriptstyle 0\leq j\leq N-1}{j\neq a}}}
\newcommand{\co}{{\cal O}}
\newcommand{\ch}{\mbox{ch}}
\newcommand{\td}{\mbox{Td}}
\newcommand{\Kappa}{\mbox{\Large $\kappa$}}
\begin{document}

\begin{titlepage}
\nopagebreak
\begin{flushright}
KUCP-167\\
hep-th/0003166
\end{flushright}

\vfill
\begin{center}
{\LARGE Comments on Central Charge }

~

{\LARGE of Topological Sigma Model }

~

{\LARGE   with Calabi-Yau Target Space }

\vskip 12mm

{\large Katsuyuki~Sugiyama${}^{\dag}$}

\vskip 10mm
{\sl Department of Fundamental Sciences}\\
{\sl Faculty of Integrated Human Studies, Kyoto University}\\
{\sl Yoshida-Nihon-Matsu cho, Sakyo-ku, Kyoto 606-8501, Japan}\\
\vspace{1cm}

     \begin{large} ABSTRACT \end{large}
\par
\end{center}
\begin{quote}
 \begin{normalsize}

We study a central charge $Z$ of a one parameter family of Calabi-Yau 
$d$-fold embedded in $CP^{d+1}$.
For a $d$-fold case, we construct the $Z$ concretely and 
analyze charge vectors of D-branes and intersection forms of 
associated cycles. We find the charges are described as 
some kinds of Mukai vectors. They are represented as products of 
Chern characters of coherent sheaves restricted on the Calabi-Yau 
hypersurfaces and square roots of A-roof genera of the $d$-folds.
By combining results of the topological sigma model and 
the data of the CFT calculations in the Gepner model, 
we find that the $Z$ is determined and is specified by a set of
  integers.
It labels boundary states in special classes where associated states 
are represented 
as tensor products of boundary states for constituent minimal models.
The $Z$ has a moduli parameter ``$t$'' that describes a deformation of a 
moduli space in 
the open string channel with  
B-type boundary conditions. 
Also monodromy matrices and homology cycles
are investigated. 

\end{normalsize}
\end{quote}

\mbox{}\\
\mbox{}\hspace{21mm}
\vfill 
\noindent
\rule{7cm}{0.5pt}\par
\vskip 1mm
{\small \noindent ${}^{\dag}$  
E-mail :\tt{} sugiyama@phys.h.kyoto-u.ac.jp}
\end{titlepage}
\vfill

\section{Introduction}
D-branes play important roles to describe solitonic modes in 
string theory. The physical observables of D-brane's 
effective theories have dependences of moduli on 
compactified internal spaces or wrapped D-branes. In this paper, we 
focus on the central charge of the type II superstring compactified on 
Calabi-Yau manifold and study its properties from the point of view of 
topological sigma models (A- and B-models)\cite{W}-\cite{KS8}. 
The central charge is 
characterized by D-brane's charges and periods of the B-model in the 
closed string channel. Together with the {\kae} potential $K$, it allows 
us to construct a BPS mass formula of D-branes wrapped around cycles. 

Recently there is a great advance \cite{Douglas} 
to study properties of charges,
boundary states\cite{Ishi},\cite{Cardy}
based on the Gepner model\cite{Gepner} associated with the 
CY${}_3$. Also there appear many consistency checks about these charges, 
intersection forms of homology cycles by analyses both in the CFT and
in the sigma models\cite{bCFT}. 
They investigated three dimensional Calabi-Yau cases. 

For these three fold cases, one can easily
construct associated canonical bases of period integrals because  
there are prepotentials for the models. It makes analyses for three folds 
comparatively tractable. 
But for $d$-fold cases $(d>3)$, we do not know any existence of
analogs of these convenient prepotentials and we cannot apply 
analogous recipes to $d$-fold cases $(d>3)$ directly. 
The aim of this paper is to develop a method to construct central
charges $Z$ of Calabi-Yau $d$-folds and to investigate their properties
in order to understand structures of the moduli spaces in the open string 
channel. We interpret constituent elements of the $Z$ as topological 
objects from the point of view of topological sigma models.

The paper is organized as follows. In section 2, we explain a mirror
manifold paired with a Calabi-Yau $d$-fold embedded in $CP^{d+1}$. 
We also explain the results in \cite{KS7},\cite{KS8} about a {\kae}
potential in order to fix notations. There we introduce a few sets of 
periods applicable either in the small or large complex structure regions.
We construct a formula of the central charge $Z$ by using the periods.
In section 3, we investigate the results by P.~Candelas et al
\cite{CDGP} in our bases and develop a method to construct the $Z$. 
By generalizing a consideration in the quintic case, we apply the method 
to the $5$-fold case and construct the central charge $Z$ concretely in 
section 4. Also the monodromy matrices associated with singular points are
investigated. In section 5, we review the Gepner model \cite{Gepner}
shortly and analyze D-brane's charges both in large radius basis and 
in the Gepner basis with B-type boundary conditions. 
In section 6, cycles associated with the periods are constructed 
explicitly and intersection forms of the cycles are studied. 
In section 7, we propose formulae of 
the $Z$ applicable in the large volume
region and make a consideration about an analogous structure to the 
Mukai vector in the $Z$. 
It is interpreted as a product of Chern characters of 
sheaves and a square root of the Todd class (or the A-roof genus) 
of the $d$-fold $M$. 
The set of sheaves is constructed as 
a dual basis of tautological line bundles of the 
ambient space $CP^{N-1}$ by a restriction on $M$.
Section 8 is devoted to conclusions and
comments. In appendix A, we summarize several examples of the 
$\sqrt{\hat{A}}$ in lower dimensional cases. In appendix B, we collect 
several data about monodromy properties of the quintic.

\section{Periods and K\"ahler Potential}

In our previous papers\cite{KS7},\cite{KS8}, 
we determine the formula of the {\kae} potential 
of the Calabi-Yau $d$-fold embedded in $CP^{d+1}$ 
\ba
M\,;\,p=X^N_1+X^N_2+\cdots +X^N_N-N\psi X_1X_2\cdots X_N=0\,,\nom
\ea
and that of the mirror partner
with a moduli parameter $\psi$ of the complex structure
\ba
W\,;\,\widehat{\{p=0\}/{\bf Z}_N^{\otimes (N-1)}}\,.
\ea
The $N$ is related with the complex dimension $d$ of M, $N=d+2$.
The $G={\bf Z}_N^{\otimes (N-1)}$ is a maximally discrete
group of the $M$.
The formula is constructed by requiring 
consistency conditions  with the results of the CFT at 
the Gepner point. In this paper, we investigate 
the central charge of the topological A-model associated with the $M$
(equivalently, the open string with B-type boundary conditions).
The formula is important for BPS mass analyses of the D-branes.

First we review the results in \cite{KS7},\cite{KS8}.
When one considers the Hodge structure of the $G$-invariant parts of the 
cohomology group H${}^d(W)$, the decomposition of the 
structure is controlled by one complex moduli parameter $\psi$.
The {\kae} potential 
$K$ in the B-model moduli of $W$ 
is constructed by combining a set of periods $\tpi_k$ quadratically
\ba
&&e^{-K}=\sum_{k=1}^{N-1}I_{k}\tpi^\dagger_k\tpi_k\,,\label{KO}\non
&&I_k=\frac{1}{\pi^N\cdot N^{N+2}}(-1)^{k-1}
\left(\sin \frac{\pi k}{N}\right)^N\,,\non
&&\tpi_k(\psi)
=\left[\Gamma\!\left(\frac{k}{N}\right)\right]^N
\frac{(N\psi)^{k}}{\Gamma (k)}\non
&&\qquad \qquad \times\left[\sum_{n=0}^{\infty}
\left[\frac{\Gamma\!\left(\frac{k}{N}+n\right)}
{\Gamma\!\left(\frac{k}{N}\right)}\right]^N\frac{\Gamma (k)}{\Gamma (Nn+k)}
(N\psi)^{Nn}\right]\,.\label{pote}
\ea
The coefficients $I_k$ are determined in \cite{KS7}.

\subsection{Periods}

The formulae Eqs.(\ref{pote}) 
are valid in the small $\psi$ region because of the 
convergence of the series expansions. 
At a point $\psi =0$ in the B-model moduli space, there is a 
${\bf Z}_N$ symmetry which rotates the $\psi\rightarrow \alpha\psi$
($\alpha =e^{2\pi i/N}$). 
A cyclic ${\bf Z}_N$ monodromy transformation ${\cal A}$ is diagonalized 
on the set of this basis $\tpi_k$ ($1\leq k\leq N-1$) with $\alpha
=e^{2\pi i/N}$
\ba
{\cal A}\tpi_k(\psi)=\alpha^k\tpi_k(\psi)\,\,\,\,\,\,\,\,
(k=1,2,\cdots ,N-1)\,.\nom
\ea
For a later convenience, we also introduce a set of periods $\vpi_j$ 
($0\leq j\leq N-1$) as linear combinations of the $\tpi_k$s
\ba
\vpi_j=-\frac{1}{N}\frac{1}{(2\pi i)^{N-1}}
\sum_{k=1}^{N-1}\alpha^{jk}(\alpha^k-1)^{N-1}\tpi_k\,.\nom
\ea
The ${\bf Z}_N$ transformation acts on this basis cyclically
\ba
{\cal A}\vpi_j(\psi)=\vpi_{j+1}(\psi),\,\,\,\,\,\,\,
(j=0,1,2,\cdots ,N-1)\,.\nom
\ea
Here we identify the $\vpi_N$ with the $\vpi_0$. 
It is a redundant basis to represent the
monodromy transformation because a linear relation is satisfied 
\ba
\sum^{N-1}_{j=0}\vpi_j=0\,.\nom
\ea
But the basis is useful because it is directly related to the Gepner
basis of the CFT. 
These two sets  of periods $\tpi_k$ and $\vpi_j$ 
are meaningful only in the small $\psi$
region because of the convergence of the series expansions. 
In order to describe the large complex structure region of $W$, we must
introduce another set of periods $\{\Omega_{m}\}$s ($m=0,1,2,\cdots ,N-2$).
A generating function of the $\Omega_m$ is defined 
by using a formal parameter $\rho$ with $\rho^{N-1}=0$
\ba
&&\sum^{N-2}_{m=0}\Omega_m\rho^m =\sqrt{\hat{K}(\rho)}
\cdot \varpi \left(\frac{\rho}{2\pi i};z\right)\,,\non
&&\vpi (v)=z^v\cdot \sum_{n=0}^{\infty}
\frac{a(n+v)}{a(v)}z^n\,,\,\,\,z=(N\psi)^{-N}\,,\label{series}\\
&&a(v)=\frac{\Gamma\!(Nv+1)}{[\Gamma\!(v+1)]^N}\,,\non
&&\hat{K}(\rho):=
\exp\left[2\sum_{m=1}\frac{N-N^{2m+1}}{2m+1}\zeta (2m+1)
\left(\frac{\rho}{2\pi i}\right)^{2m+1}
\right]\non
&&\qquad =1+2\zeta (3)\frac{c_3}{N}\left(\frac{\rho}{2\pi i}\right)^3+
{\cal O}(\rho^5)\,.\nom
\ea
The infinite series Eq.(\ref{series}) converges around the 
large complex structure point $z\sim 0$ of W.
We find that the two sets of the periods $\tpi_k$ and $\Omega_{\ell}$ 
are related by a transformation 
matrix $\tilde{M}$ with components $\tilde{M}_{k\ell}$
through an analytic continuation into the large complex structure region
\ba
&&\tpi_k=\sum^{N-2}_{\ell =0}\tilde{M}_{k\ell}\Omega_{\ell}\,\,\,\,\,
(k=1,2,\cdots ,N-1)\,,\non
&&\tilde{M}_{k\ell}=(-N)\cdot (2\pi i)^{N-1}
\times \left[
 \sqrt{\hat{A}(\rho)}\cdot 
\frac{\alpha^k}{e^{\rho}-\alpha^k}\cdot (-\rho)^{\ell}
\right]\Biggr|_{\rho^{N-2}}\non
&&\qquad =
(-N)\cdot 
(2\pi i)^{N-1}\cdot \sum^{N-2}_{m=0 }
G_{k,m}V_{m,\ell}\,,\non 
&&G_{k,m}=\frac{-\alpha^k}{(\alpha^k -1)^{m+1}}\,\,\,\,\,
(1\leq k\leq N-1\,,\,\,0\leq m \leq N-2)\,,\non
&&V_{m,\ell}=\left[
\sqrt{\hat{A}(\rho)}\cdot (e^{\rho}-1)^m \cdot (-\rho)^{\ell}
\right]\Biggr|_{\rho^{N-2}}
\,\,\,\,\,(0\leq m\leq N-2\,,\,\, 0\leq \ell \leq N-2)\,.\label{coeff1}
\ea
The transformation matrix $V$ contains
a square root of a topological invariant ``A-roof genus'' 
of the Calabi-Yau space
\ba
&&\hat{A}(\rho)=
\left(\frac{\dis \frac{\rho}{2}}{\dis \sinh \frac{\rho}{2}}\right)^N
\cdot\left(\frac{\dis \sinh \frac{N\rho}{2}}{\dis \frac{N\rho}{2}}\right)\non
&&\qquad 
=\exp\left[+\sum_{m=1}\frac{(-1)^mB_m}{(2m)!}\frac{N-N^{2m}}{2m}\rho^{2m}
\right]\non
&&\qquad =1+\frac{1}{12}\frac{c_2}{N}\rho^2 +{\cal O}(\rho^4)\,.\nom
\ea
The $B_m$s are Bernoulli numbers and are defined in our convention as
\ba
&&\frac{x}{e^x-1}=1-\frac{x}{2}-\sum^{\infty}_{n=1}
\frac{(-1)^n\cdot B_n}{(2n)!}x^{2n}\,,\non
&&B_1=\frac{1}{6}\,,\,\,B_2=\frac{1}{30}\,,\,\,
B_3=\frac{1}{42}\,,\,\,B_4=\frac{1}{30}\,,\cdots\,\,.\nom
\ea
The expansion coefficients of the $\hat{A}(\rho)$ in terms of the $\rho$
are represented as some combinations of Chern classes of the $M$. 
We summarize several concrete examples of the $\sqrt{\hat{A}(\rho)}$ in
the appendix A.
Also we define auxiliary sets of periods $\{\hat{\Pi}_n\}$ and  
$\{\hat{\hat{\Pi}}_m\}$
($n,m=0,1,2,\cdots ,N-2$)
\ba
&&\hat{\Pi}_n=\sum_{\ell =0}^{N-2}V_{n,\ell}\Omega_{\ell}\,,\,\,\,
\hat{\hat{\Pi}}_m=\sum_{n =0}^{N-2}U_{m,n}\Pi_n\,,\non
&&V_{n,\ell}=\left[\sqrt{\hat{A}(\rho)}\cdot (e^{\rho}-1)^n
\cdot (-\rho)^{\ell}\right]_{\rho^{N-2}}\,,\,\,\,
U_{m,n}=\left(\matrix{m\cr n}\right)\cdot (-1)^{m-n}\,.\nom
\ea
Here we introduced a symbol $\dis \left(\matrix{a\cr b}\right)$ as
a ratio of Euler's gamma functions
\ba
\left(\matrix{a\cr b}\right):=\frac{\Gam (a+1)}{\Gam (b+1)\Gam (a-b+1)}\,.\nom
\ea
The $\hat{\Pi}$ is related to the basis $\vpi_j$ through an analytic
continuation 
\ba
&&\varpi_j=\sum_{n=0}^{N-2}P_{j,n}\hat{\Pi}_n\,,\non
&&P_{j,n}=\delta_{n,N-2}-N\cdot \left(\matrix{N-2-n\cr
j-1-n}\right)\times
(-1)^{j-1-n}\,,\,\,\,
\left(\matrix{j=0,1,\cdots ,N-1\cr n=0,1,\cdots ,N-2}\right)\,.\label{coeff2}
\ea
In discussing properties in the large $\psi$ region, we use the sets 
$\hat{\Pi}_{n}$ or $\Omega_m$.

\subsection{K\"ahler Potential}

Under this preparation, we shall write down our results in the {\kae}
potential $K$ in these bases
\ba
&&e^{-K}=\sum^{N-1}_{\ell =1}
\tilde{\varpi}_{\ell}^\dagger I_{\ell}
\tilde{\varpi}_{\ell}\non
&&\qquad =\frac{(-2\pi i)^{N-2}}{N^{N+2}}\cdot \sum_{j\neq a}
\sum_{j'\neq a}{\varpi}_{j}^\dagger {\cal K}_{j,j'}
{\varpi}_{j'}\non
&&\qquad =\frac{-1}{(2\pi iN)^N}\cdot\frac{1}{N^2}\cdot \sum_{m,n=0}^{N-2}
\hat{\Pi}_{m}^\dagger {\cal I}_{m,n}
\hat{\Pi}_{n}\,,\non
&&\qquad =(-1)^N\left(\frac{2 \pi i}{N}\right)^{N-2}\cdot 
\frac{1}{N^2}\sum_{\ell ,\ell'=0}^{N-2}\Omega^{\dagger}_{\ell}
\Sigma_{\ell ,\ell'}\Omega_{\ell'}\,,\non
&&I_{\ell}=\frac{(-1)^{\ell -1}}{\pi^N\cdot N^{N+2}}\cdot 
\left(\sin \frac{\pi\ell}{N}\right)^N
\,,\,\,\,\,\,\,(\ell =1,2,\cdots ,N-1)\,,\non
&&{\cal K}_{j,j'}=\frac{i^N}{2^{N-2}}\times 
\sum^{N-1}_{\ell =1}(-1)^{\ell}\left(\sin \frac{\pi\ell}{N}
\right)^{-(N-2)}\times 
(\alpha^{j\ell}-\alpha^{a\ell})(\alpha^{-j'\ell}-\alpha^{-a\ell})\,,
\,\,\,\,\,\,\,\non
&&\hspace{10cm} (j,j'=0,1,2,\cdots,N-1)\,,\non
&&{\cal I}_{m,n}=2^N\cdot i^N\cdot \sum^{N-1}_{k=1}
\frac{(-1)^k\cdot\left(\dis \sin \frac{\pi k}{N}\right)^N }
{(\alpha^{-k}-1)^{m+1}(\alpha^{k}-1)^{n+1}}\,,\,\,\,\,\,\,\,
(m,n=0,1,2,\cdots,N-2)\,,\non
&&\Sigma_{\ell ,\ell'}=(-1)^{\ell}\cdot \delta_{\ell +\ell' ,N-2}\,,
\,\,\,\,\,\,\,
(\ell ,\ell'=0,1,2,\cdots,N-2)\,.\nom
\ea
In the above formulae, the $I_k$, ${\cal K}_{j,j'}$, ${\cal I}_{m,n}$
 and $\Sigma_{\ell ,\ell'}$ are intersection matrices associated with
 homology cycles for the corresponding periods $\tpi_k$, $\vpi_j$, 
$\hat{\Pi}_m$ and $\Omega_{\ell}$. 
The matrix $I$ has a diagonal form. 
In contrast, the ${\cal K}_{j,j'}$
and ${\cal I}_{m,n}$ are lower triangular matrices with 
non-vanishing components in the right lower entries
with $j+j'\geq N-2$, $m+n\geq N-2$.
Also the determinant of the ${\cal K}$ is not unit, but $\det{\cal
K}=N^2$. So the associated cyclic basis is not a canonical one.
The set of the $\Omega_{\ell}$s has an intersection matrix 
$\Sigma_{\ell,\ell'}$ with non-vanishing components at $\ell +\ell'=N-2$.
The form of the $\Sigma_{\ell ,\ell'}$
means that the set of periods $\Omega_{\ell}$s is a 
symplectic or an SO-invariant basis and associated homology cycles have 
appropriate intersection forms. 
But it is not an integral basis and the 
associated cycles belong to homology classes 
$\bigoplus^{d}_{\ell =0}\mbox{H}_{2\ell}(M\,;\,{\bf Q})$ or 
$\mbox{H}_{d}(W\,;\,{\bf Q})$ respectively in the A-, B-models.
In order to obtain a set of canonical basis $\{\Pi_m\}$ 
($m=0,1,2,\cdots ,N-2$), we have to
perform some linear transformation on the $\Omega_{\ell}$ 
\ba
\Pi_m=\sum^{N-2}_{\ell =0}{\cal N}_{m,\ell}\Omega_{\ell}
\,\,\,\,\,\,\,\,(m=0,1,2,\cdots ,N-2)\,.\nom
\ea
The transformation matrix ${\cal N}$ generally has entries 
with fractional rational numbers.
The basis $\Pi$ is 
needed to discuss the D-brane charges $Q_{\ell}$s or a central
charge $Z$ in the BPS mass formula. 

\subsection{Central Charge}

In the B-model case in the open string channel\footnote{So far we used
the  words ``A-'', ``B-''models for the closed string case. But the 
definition is exchanged when we consider properties in the open string 
channels. }, there appear even dimensional D$p$-branes  
($p=0,2,4,\cdots ,2d$) which wrap around homology cycles $\Sigma_p$ 
($p=0,2,4,\cdots ,2d$) of the Calabi-Yau $d$-fold $M$. 
The brane charge $Q_{2d-2\ell}$ associated with the D-brane is
defined by integrating a gauge field, more precisely a Mukai vector 
$v({\cal E})$ associated with a bundle (sheaf) ${\cal E}$ over the cycle
$\Sigma_{2\ell}\subset M$ ($\ell =0,1,2,\cdots ,d$)
\ba
Q_{2d-2\ell}=
\int_{\Sigma_{2\ell}}v({\cal E})
\,\,\,\,\,\,\,\,(\ell =0,1,2,\cdots ,d)\,.\nom
\ea
By combining these charges $Q_p$ and the canonical basis $\Pi_{\ell}$ 
with the B-type boundary conditions, we can construct the central charge 
$Z$
\ba
Z=\sum^d_{\ell =0}Q_{2\ell}\Pi_{\ell}\,.\nom
\ea
It is a constituent block of 
a BPS mass formula $m_{BPS}\sim  e^{+K/2}|Z|$ in the curved space.
We can represent this canonical basis $\Pi_{\ell}$ in the bases 
$\Omega_{\ell}$, $\hat{\Pi}_{\ell}$ and $\vpi_j$
\ba
&&\Pi_n=\sum^{N-2}_{\ell =0}{\cal N}_{n,\ell}\Omega_{\ell}=
\sum^{N-2}_{\ell =0}S_{n,\ell}\hat{\Pi}_{\ell}\non
&&\qquad  =\wa m_{n,j}\vpi_j
\,\,\,\,\,\,\,\,(0\leq n\leq N-2)\,.\nom
\ea
There are two linear relations among the undetermined matrices 
${\cal N}$, $S$ and $m$ with Eqs.(\ref{coeff1}),(\ref{coeff2})
\ba
&&{\cal N}_{n\ell}=\sum^{N-2}_{m=0}S_{nm}V_{m\ell}\,,\,\,\,
(0\leq n\leq N-2\,;\,0\leq \ell \leq N-2)\,,\non
&&{\cal S}_{n\ell}=\wa m_{nj}P_{j\ell}\,,\,\,\,(0\leq n\leq N-2\,;\,
0\leq \ell \leq N-2)\,.\nom
\ea
If we can determine one of these matrices ${\cal N}$, $S$ and $m$, the 
canonical basis $\Pi_{\ell}$ is fixed. 
Here the ``canonical'' condition means that the set $\{\Pi_{\ell}\}$
is a symplectic (or an SO-invariant) and an integral basis of 
monodromy transformations. 
The symplectic (or SO-invariant) condition is reduced to 
that on the ${\cal N}$ with an appropriate integer $\lambda$ 
and the matrix $\Sigma$
\ba
{\cal N}^t\cdot \Sigma \cdot {\cal N}=\lambda \Sigma\,.\nom
\ea
But generally the matrix ${\cal N}$ may have fractional components. 

\section{Quintic}

In order to exemplify our considerations, we shall study the result of
the quintic. P.~Candelas et al directly constructed
the matrix $m$\cite{CDGP}, 
which connects the two bases $\Pi_{\ell}$ and $\vpi_j$
\ba
&&\Pi:={}^t\left(\matrix{\Pi_0 &\Pi_1 & \Pi_2 &  \Pi_3}\right)\,,\non
&&\vpi:={}^t\left(\matrix{\vpi_0 &\vpi_1 & \vpi_2 &
\vpi_4}\right)\,,\non
&&m=\left(
\matrix{ 1 & 0 & 0 & 0 \cr -{\frac{2}{5}} & {\frac{2}{5}} & {\frac{1}
    {5}} & -{\frac{1}{5}} \cr {\frac{21}{5}} & -{\frac{1}{5}} & -{\frac{3}
     {5}} & {\frac{8}{5}} \cr 1 & -1 & 0 & 0 \cr  } 
\right)\,,\,\,\,
{m}^{-1}:=\left(
\matrix{ 1 & 0 & 0 & 0 \cr 1 & 0 & 0 & -1 \cr -4 & 8 & 1 & 3 \cr 
    -4 & 3 & 1 & 1 \cr  } 
\right)\,.\nom
\ea
These $\Pi_{\ell}$s are canonical periods and are described by using a
prepotential\footnote{This ``$F$'' is not a gauge field ``$F$'' on
the D-brane. I believe that there is no confusion in these notations.}
 $F$
\ba
&&F=-\frac{\kappa}{6}t^3+\frac{1}{2}at^2+bt+\frac{1}{2}c+f\,,\non
&&\kappa =5\,,\,\,
a=\frac{-11}{2}\,,\,\,b=\frac{25}{12}\,,\,\,c=\frac{\chi
\zeta (3)}{(2\pi i)^3}
\,,\,\,\chi =-200\,,\non
&&\Pi =\left(\matrix{\Pi_0\cr\Pi_1\cr\Pi_2\cr\Pi_3}\right)
=\left(\matrix{1\cr t\cr \del_tF\cr t\del_tF-2F}\right)\times \Pi_0
\,,\,\,\label{quin1}\\
&&2\pi i t
=\log z+\frac{\dis\sum^{\infty}_{n=1}\frac{(5n)!}{(n!)^5}
\left(\sum^{5n}_{m=n+1}\frac{5}{m}\right)z^n}
{\dis \sum^{\infty}_{n=0}\frac{(5n)!}{(n!)^5}z^n}\,,\,\,\,
q=e^{2\pi i t}\,.\nom
\ea
The prepotential $F$ of $M$ is expressed as a sum of a polynomial part
of $t$ and a non-perturbative part $f$.
We choose a set of the $\Omega_{\ell}$s in the
$N=5$ case
\ba
&&\vpi\left(\frac{\rho}{2\pi i}\right)\sqrt{\hat{K}(\rho)}=:
\sum_{\ell \geq 0}\rho^{\ell}\Omega_{\ell}\,,\non
&&\sqrt{\hat{A}(\rho)}=1+\frac{5}{12}\rho^2\,,\,\,\,
\sqrt{\hat{K}(\rho)}=1-\frac{40}{(2\pi i)^3}\zeta (3)\rho^3\,,\non
&&\hat{c}=\frac{40}{(2\pi i)^3}\zeta (3)=-\frac{1}{5}c\,,\non
&&\Omega =\left(\matrix{\Omega_0\cr\Omega_1\cr\Omega_2\cr\Omega_3}\right)
=\left(\matrix{1\cr t\cr \frac{1}{2}t^2 +S_2(0,x_2)
\cr\frac{1}{6}t^3-\hat{c}+tS_2(0,x_2)+S_3(0,x_2,x_3)
}\right) \times \Omega_0 \,,\,\label{quin2}\\
&&x_n=\frac{1}{(2\pi i)^n}\frac{1}{n!}\del^n_{\rho }\log
\left[\sum^{\infty}_{m=0}\frac{\Gam\!(N(m+\rho)+1)}
{\Gam\!(N\rho +1)}\left(\frac{\Gam\!(\rho +1)}
{\Gam\!(m+\rho +1)}\right)^N z^m\right]\Biggr|_{\rho =0}\,.\nom
\ea
By comparing Eq.(\ref{quin1}) and Eq.(\ref{quin2}),
we can obtain a matrix ${\cal N}$  
\ba
&&\Pi ={\cal N}\Omega\,,\,\,\,
{\cal N}
=\left(
\matrix{ 1 & 0 & 0 & 0 \cr 0 & 1 & 0 & 0 \cr {\frac{25}{12}} & -{\frac{11}
     {2}} & -5 & 0 \cr 0 & -{\frac{25}{12}} & 0 & -5 \cr  } 
\right)
\,.\label{n}
\ea
The ${\cal N}$ is a kind of a symplectic matrix and satisfies a relation
\ba
{\cal N}^t\cdot \Sigma \cdot {\cal N}=(-5)\cdot\Sigma \,.\nom
\ea
Also one can check this ${\cal N}$ and the $m$ satisfy an equation
\ba
{\cal N}=m\cdot P\cdot V\,,\nom
\ea
by using definitions of the $V$ and $P$ in Eqs.(\ref{coeff1}),(\ref{coeff2})
\ba
P\cdot V=\left(
\matrix{ 1 & 0 & 0 & 0 \cr 1 & {\frac{25}{12}} & 0 & 5 \cr -{\frac{23}
     {12}} & -{\frac{15}{4}} & -5 & -15 \cr -{\frac{23}{12}} & -{\frac{55}
     {12}} & -5 & -5 \cr  } 
\right)\,.\nom
\ea
In the $\hat{\Pi}$ basis, this result is also expressed as
\ba
\Pi =S\cdot \hat{\Pi}\,,\,\,\,
S=\left(
\matrix{ 0 & 0 & 0 & 1 \cr 0 & 0 & -1 & 1 \cr 0 & -5 & 8 & 
    -3 \cr 5 & 0 & 0 & 0 \cr  } \right)=m\cdot P\,.\nom
\ea
Then the intersection form in the $\hat{\Pi}$ basis is obtained as
\ba
&&S^t\cdot \Sigma \cdot S=\left(
\matrix{ 0 & 0 & 0 & 5 \cr 0 & 0 & -5 & 5 \cr 0 & 5 & 0 & -5 \cr -5 & 
    -5 & 5 & 0 \cr  } \right)\,,\,\,\,
\Sigma =\left(\matrix{ 0 & 0 & 0 & -1 \cr 0 & 0 & 1 & 0 \cr 0 & 
    -1 & 0 & 0 \cr 1 & 0 & 0 & 0 \cr  } \right)\,.\nom
\ea
It coincides with the matrix ${\cal I}_{m,n}$ in our result for the $N=5$
case
\ba
&&S^t\cdot \Sigma \cdot S=5\cdot {\cal I}\,,\,\,\,
{\cal I}=\left(
\matrix{ 0 & 0 & 0 & 1 \cr 0 & 0 & -1 & 1 \cr 0 & 1 & 0 & -1 \cr -1 & 
    -1 & 1 & 0 \cr  } \right)\,.\,\,\,\nom
\ea
Conversely the result means that the matrix ${\cal I}$ can be
factorized into a form 
\ba
{\cal I}=\frac{1}{5}S^t\cdot \Sigma \cdot S\,,\nom
\ea
with a triangular matrix $S$ with non-vanishing components in the 
right lower entries. 
Generally we know the forms of the matrices $V$ and $P$ 
in Eqs.(\ref{coeff1}),(\ref{coeff2}), 
but the remaining ones $m$ and ${\cal N}$ 
are unknown. If we can determine the ${\cal N}$, then we obtain the $m$
from a relation
\ba
m={\cal N}\cdot (P\cdot V)^{-1}\,.\label{relation}
\ea
This ${\cal N}$ may contain invariants of K(M)-theory of the D-branes as
its components. 
The ${\cal N}$ is a triangular matrix with non-vanishing components in
the left lower blocks 
\ba
{\cal N}=\left(\matrix{*&0&0&0\cr *&*&0&0\cr *&*&*&0\cr *&*&*&*}\right)\,.\nom
\ea
We multiply 
a charge vector $Q$ on the ${\cal N}$ from the left. The $Q$ encodes
information of homology cycles of the Calabi-Yau around which 
D-branes wrap. 
For the quintic case, numbers in 
the last row of the ${\cal N}$ in 
Eq.(\ref{n}) coincide with coefficients of a topological 
invariant $(-5)\sqrt{\hat{A}(\rho)}$ 
\ba
(-5)\sqrt{\hat{A}(\rho)}=-5-\frac{25}{12}\rho^2\,.\nom
\ea
In particular, the $25/12$ is related with a $2$nd Chern number $c_2=50$ 
of $M$ 
\ba
\frac{1}{24}c_2=\frac{25}{12}\,.\nom
\ea
It also appears in the $1$st column in the ${\cal N}$. 
The number $5$ is interpreted as a triple intersection number of a 
$4$-cycle in the $M$.
The remaining entry $-11/2$ is expected to be interpreted as 
some kind of invariant associated with a normal bundle of 
a world volume of a D-brane. But we do not precisely know the geometric 
characterization of this yet and cannot explain this number 
from the point of view of characteristic classes of some associated
bundle.

We will return to the central charge $Z$. 
If we can find the matrix ${\cal N}$ and choose the basis $\Pi$, 
the central charge is evaluated in the B-model in the open string channel 
\ba
&&Z=Q\cdot \Pi =(Q{\cal N})({\cal N}^{-1}\Pi)
=(Q{\cal N})\cdot\Omega \,,\non
&&Q=\left(\matrix{Q_{2d}& Q_{2d-2}& \cdots & Q_2& Q_0}\right)\,.\nom
\ea
When one uses a basis $\Omega_{\ell}$ to construct the $Z$, 
the formula of a modified charge vector 
$Q\cdot {\cal N}$ is needed. But no systematic method to calculate the
${\cal N}$ is known yet. Also the basis $\Pi$ has an ambiguity in 
the multiplication with a matrix $L$ from the left-side
\ba
\Pi'=L\cdot \Pi =L\cdot {\cal N}\Omega\,.\nom
\ea
When the $L$ is a symplectic (or an SO-invariant) 
matrix with integer components
\ba
L^t\cdot \Sigma \cdot L=\Sigma\,,\nom
\ea
the modified $\Pi'$ gives us the same {\kae} potential as that in the
$\Pi$ basis. It leads to the same results about properties of 
closed string moduli spaces. We do not know any principle to fix the 
ambiguity completely and do not touch on this here. 
In this paper, we choose a standard convention $L=I$ by choosing a pure
D${}_{2d}$ charge in the next section.

\section{$5$-Fold}

In the previous section, we analyzed the quintic. 
We started analyses by using the prepotential $F$
of the quintic and studied its monodromy properties.
We cannot expect to use analogs of prepotentials for other 
$d$-fold cases $(d>3)$.
But the essential part we learned in the previous section  
seems to be a factorizable property of the matrix ${\cal I}$ by 
a (triangular) matrix $S$ and a matrix $\Sigma$.
Under this consideration, we will investigate the $N=7$ ($d=5$) case
concretely. First the ${\cal I}$ in the $\hat{\Pi}$ basis is defined as
\ba
&&{\cal I}=\left(
\matrix{ 0 & 0 & 0 & 0 & 0 & 1 \cr 0 & 0 & 0 & 0 & 
    -1 & 2 \cr 0 & 0 & 0 & 1 & -1 & -1 \cr 0 & 0 & -1 & 0 & 2 & 
    -2 \cr 0 & 1 & 1 & -2 & 0 & 3 \cr -1 & -2 & 1 & 2 & -3 & 0 \cr  } 
\right)\,.\nom
\ea
We find that the ${\cal I}$ is factorized into a form with a matrix
$\Sigma$ and a triangular matrix $S$
\ba
&&7\cdot {\cal I}=S^t\cdot\Sigma \cdot S\,,\non
&&\Sigma =\left(
\matrix{ 0 & 0 & 0 & 0 & 0 & -1 \cr 0 & 0 & 0 & 0 & 1 & 0 \cr 0 & 0 & 0 & 
    -1 & 0 & 0 \cr 0 & 0 & 1 & 0 & 0 & 0 \cr 0 & 
    -1 & 0 & 0 & 0 & 0 \cr 1 & 0 & 0 & 0 & 0 & 0 \cr  } \right)\,,\,\,\,
S=\left(
\matrix{ 0 & 0 & 0 & 0 & 0 & 1 \cr 0 & 0 & 0 & 0 & -1 & 2 \cr 0 & 0 & 0 & 
    -1 & 1 & 1 \cr 0 & 0 & -7 & 7 & -7 & 7 \cr 0 & -7 & 0 & 14 & 
    -21 & 7 \cr 7 & 0 & 0 & 0 & 0 & 0 \cr  } 
\right)\,.\nom
\ea
This decomposition is not unique as we discussed at the end of the 
previous section. There are some possibilities for 
choosing the $S$.
When we pick a cyclic basis  $\{\vpi_j\}$ with an index ``$j$'' as
\ba
\vpi={}^t
\left(\matrix{\vpi_0 &\vpi_1 &\vpi_2 &\vpi_3 &\vpi_5 &\vpi_6 }\right)\,,\nom
\ea
a transformation matrix $P$ from the $\hat{\Pi}$ to the $\vpi$ 
$(\hat{\Pi}=P\cdot \vpi)$ 
is 
obtained
\ba
&&{P}=\left(
\matrix{ 0 & 0 & 0 & 0 & 0 & 1 \cr -7 & 0 & 0 & 0 & 0 & 1 \cr 35 & 
    -7 & 0 & 0 & 0 & 1 \cr -70 & 28 & -7 & 0 & 0 & 1 \cr -35 & 28 & 
    -21 & 14 & -7 & 1 \cr 7 & -7 & 7 & -7 & 7 & -6 \cr  } 
\right)\,.\nom
\ea
Now we know the matrix $S$, we can determine matrices $m$ and ${\cal N}$.
By using the relation 
$S=m\cdot P$, we can obtain a transformation matrix $m$ from the 
$\vpi$ basis to the $\Pi$ basis
\ba
&&\Pi ={}^t
\left(\matrix{\Pi_0 &\Pi_1 &\Pi_2 &\Pi_3 &\Pi_4 &\Pi_5  }\right)\,,\non
&&m=\left(
\matrix{ 1 & 0 & 0 & 0 & 0 & 0 \cr -{\frac{3}{7}} & {\frac{3}{7}} & {\frac{2}
    {7}} & {\frac{1}{7}} & -{\frac{1}{7}} & -{\frac{2}{7}} \cr {\frac{3}
    {7}} & {\frac{6}{7}} & {\frac{3}{7}} & {\frac{1}{7}} & 0 & {\frac{1}
    {7}} \cr -4 & 4 & 1 & 0 & 0 & -1 \cr -8 & -4 & -3 & -1 & -1 & 
    -4 \cr 1 & -1 & 0 & 0 & 0 & 0 \cr  } 
\right)\,.\label{m}
\ea
Characteristic features of the $m$ appear in the $1$st and the last rows in 
$m$. The first row means that the $\Pi_0$ is identified with the 
$\vpi_0$. 
The last row implies that the $\Pi_5$ is represented as 
a linear combination of only $\vpi_0$ and $\vpi_1$ as
\ba
\Pi_5 =\vpi_0-\vpi_1\,.\nom
\ea
We believe that these structures are universal. 
Especially the $\Pi_{d}$ might be represented as
\ba
\Pi_d=\vpi_0-\vpi_1\,.\nom
\ea
It is related to the structure of a pure D${}_{2d}$-brane charge $Q_{2d}$.

Also we find the matrix ${\cal N}$ for the $N=7$ case
by using an equation Eq.(\ref{relation})
\ba
&&\Pi ={\cal N}\Omega\,,\non
&&{\cal N}=\left(
\matrix{ 1 & 0 & 0 & 0 & 0 & 0 \cr 0 & 1 & 0 & 0 & 0 & 0 \cr {\frac{7}
    {8}} & {\frac{1}{2}} & -1 & 0 & 0 & 0 \cr 0 & {\frac{161}
    {24}} & 0 & 7 & 0 & 0 \cr -{\frac{11711}{1920}} & {\frac{161}{48}} & 
    {\frac{161}{24}} & {\frac{7}{2}} & -7 & 0 \cr 0 & {\frac{147}
    {640}} & 0 & -{\frac{49}{8}} & 0 & -7 \cr  } 
\right)\,,\non
&&{\cal N}^t\cdot \Sigma \cdot {\cal N}=(-7)\cdot \Sigma\,.\nom
\ea
The entries in the last row of the ${\cal N}$ are topological numbers. 
They coincide with coefficients of the $(-7)\sqrt{\hat{A}(\rho)}$ for
the $N=7$ case
\ba
(-7)\sqrt{\hat{A}(\rho)}=-7-\frac{49}{8}\rho^2 +\frac{147}{640}\rho^4\,.\nom
\ea
The entries $\pm 7$ is associated with an intersection number $7$ of 
five $8$-cycles in $M$.
Possibly the other entries could be interpreted as invariants associated 
with D-branes from the point of view of K(M)-theory. 
We cannot precisely know any geometric characterization 
of numbers at the other entries yet. We do not touch on this topic
here. 

Next let us consider monodromy matrices associated with singular points
in order to confirm our result for the $N=7$ case. The $5$-fold has 
singular points $\psi=0,\infty ,e^{2\pi i\ell/7}$ 
($\ell =0,1,2\cdots ,6$) in the B-model moduli space. 
At the point $\psi =0$, the monodromy transformation acts on the basis 
$\vpi$ cyclically and is realized as a matrix $A_{\vpi}$
\ba
{\cal A}\vpi =A_{\vpi}\vpi\,.\nom
\ea
The action of ${\cal A}$ on other bases $\hat{\Pi}$, $\Pi$ and $\Omega$ 
is calculated as representation 
matrices $A_{\hat{\Pi}}$, $A_{\Pi}$ and $A_{\Omega}$ 
by using the ${\cal N}$ and $m$ 
\ba
&&A_{\varpi}=\left(
\matrix{
   0 & 1 & 0 & 0 & 0 & 0 \cr 0 & 0 & 1 & 0 & 0 & 0 \cr 0 & 0 & 0 & 1 & 0 & 
   0 \cr -1 & -1 & -1 & -1 & -1 & 
    -1 \cr 0 & 0 & 0 & 0 & 0 & 1 \cr 1 & 0 & 0 & 0 & 0 & 0 \cr  } 
\right)\,,\,\,\,
A_{\hat{\Pi}}=\left(
\matrix{ -6 & 1 & 0 & 0 & 0 & 0 \cr -21 & 1 & 1 & 0 & 0 & 0 \cr 
    -35 & 0 & 1 & 1 & 0 & 0 \cr -35 & 0 & 0 & 1 & 1 & 0 \cr 
    -21 & 0 & 0 & 0 & 1 & 1 \cr -7 & 0 & 0 & 0 & 0 & 1 \cr  } 
\right)\,,\non
&&A_{\Pi}=\left(
\matrix{ 1 & 0 & 0 & 0 & 0 & -1 \cr -1 & 1 & 0 & 0 & 0 & 1 \cr 
    -1 & 1 & 1 & 0 & 0 & 1 \cr -14 & 0 & 7 & 1 & 0 & 14 \cr -7 & 
    -14 & 7 & 1 & 1 & 7 \cr 7 & 7 & -14 & 0 & -1 & -6 \cr  } 
\right),\,
A_{\Omega}=\left(
\matrix{ 1 & -{\frac{147}{640}} & 0 & {\frac{49}{8}} & 0 & 7 \cr -1 & {
     \frac{787}{640}} & 0 & -{\frac{49}{8}} & 0 & -7 \cr {\frac{1}{2}} & 
    -{\frac{6737}{5120}} & 1 & {\frac{539}{64}} & 0 & {\frac{77}{8}} \cr 
    -{\frac{1}{6}} & {\frac{757}{1024}} & -1 & -{\frac{1033}{192}} & 0 & 
    -{\frac{175}{24}} \cr {\frac{1}{24}} & -{\frac{330779}{1228800}} & {
     \frac{1}{2}} & {\frac{26633}{15360}} & 1 & {\frac{5999}{1920}} \cr -
     {\frac{1}{120}} & {\frac{85451}{1228800}} & -{\frac{1}{6}} & -{\frac{
       3737}{15360}} & -1 & {\frac{289}{1920}} \cr  } \right).\nom
\ea
Also the transformation around the $\psi =\infty$ point is evaluated on 
these bases as matrices $T_{\vpi}$, $T_{\hat{\Pi}}$, $T_{\Pi}$ and
$T_{\Omega}$
\ba
&&T_{\varpi}=\left(
\matrix{ 1 & 0 & 0 & 0 & 0 & 0 \cr 2 & 0 & 0 & 0 & 0 & -1 \cr 
    -6 & 1 & 0 & 0 & 0 & 6 \cr 15 & 0 & 1 & 0 & 0 & -15 \cr 14 & -1 & -1 & 
    -1 & -1 & -16 \cr -6 & 0 & 0 & 0 & 1 & 6 \cr  } 
\right)\,,\,\,\,
T_{\hat{\Pi}}=\left(
\matrix{ 1 & -1 & 1 & -1 & 1 & -1 \cr 0 & 1 & -1 & 1 & 
    -1 & 1 \cr 0 & 0 & 1 & -1 & 1 & -1 \cr 0 & 0 & 0 & 1 & 
    -1 & 1 \cr 0 & 0 & 0 & 0 & 1 & -1 \cr 0 & 0 & 0 & 0 & 0 & 1 \cr  } 
\right)\,,\non
&&T_{\Pi}=\left(
\matrix{ 1 & 0 & 0 & 0 & 0 & 0 \cr 1 & 1 & 0 & 0 & 0 & 0 \cr 0 & 
    -1 & 1 & 0 & 0 & 0 \cr 14 & 7 & -7 & 1 & 0 & 0 \cr 7 & 14 & 0 & 
    -1 & 1 & 0 \cr -7 & -7 & 14 & -1 & 1 & 1 \cr  } 
\right)\,,\,\,\,
T_{\Omega}=\left(
\matrix{ 1 & 0 & 0 & 0 & 0 & 0 \cr 1 & 1 & 0 & 0 & 0 & 0 \cr {\frac{1}
    {2}} & 1 & 1 & 0 & 0 & 0 \cr {\frac{1}{6}} & {\frac{1}
    {2}} & 1 & 1 & 0 & 0 \cr {\frac{1}{24}} & {\frac{1}{6}} & {\frac{1}
    {2}} & 1 & 1 & 0 \cr {\frac{1}{120}} & {\frac{1}{24}} & {\frac{1}
    {6}} & {\frac{1}{2}} & 1 & 1 \cr  } \right)\,.\nom
\ea
The forms of the matrices $T_{\hat{\Pi}}$ and $T_{\Omega}$ are
universal. We will explain these points in the section \ref{radius}.

The $\psi =1$ is a conifold-like point and 
associated monodromy matrices are obtained
for these bases as $P_{\vpi}$, $P_{\Pi}$, $P_{\hat{\Pi}}$ and $P_{\Omega}$
\ba
&&P_{\varpi}=\left(
\matrix{ 2 & -1 & 0 & 0 & 0 & 0 \cr 1 & 0 & 0 & 0 & 0 & 0 \cr 
    -6 & 6 & 1 & 0 & 0 & 0 \cr 15 & -15 & 0 & 1 & 0 & 0 \cr 15 & 
    -15 & 0 & 0 & 1 & 0 \cr -6 & 6 & 0 & 0 & 0 & 1 \cr  } 
\right)\,,\,\,\,
P_{\hat{\Pi}}=\left(
\matrix{
   1 & 0 & 0 & 0 & 0 & 0 \cr 7 & 1 & 0 & 0 & 0 & 0 \cr 14 & 0 & 1 & 0 & 0 & 
   0 \cr 21 & 0 & 0 & 1 & 0 & 0 \cr 14 & 0 & 0 & 0 & 1 & 0 \cr 7 & 0 & 0 & 
   0 & 0 & 1 \cr  } 
\right)\,,\non
&&P_{\Pi}=\left(
\matrix{
   1 & 0 & 0 & 0 & 0 & 1 \cr 0 & 1 & 0 & 0 & 0 & 0 \cr 0 & 0 & 1 & 0 & 0 & 
   0 \cr 0 & 0 & 0 & 1 & 0 & 0 \cr 0 & 0 & 0 & 0 & 1 & 0 \cr 0 & 0 & 0 & 0 & 
   0 & 1 \cr  } 
\right)\,,\,\,\,
P_{\Omega}=\left(
\matrix{ 1 & {\frac{147}{640}} & 0 & -{\frac{49}{8}} & 0 & 
    -7 \cr 0 & 1 & 0 & 0 & 0 & 0 \cr 0 & {\frac{1029}{5120}} & 1 & -{
      \frac{343}{64}} & 0 & -{\frac{49}
     {8}} \cr 0 & 0 & 0 & 1 & 0 & 0 \cr 0 & -{\frac{3087}{409600}} & 0 & 
    {\frac{1029}{5120}} & 1 & {\frac{147}
    {640}} \cr 0 & 0 & 0 & 0 & 0 & 1 \cr  } \right)\,.\nom
\ea
Under these monodromy transformations, the set $\Omega_{\ell}$ is 
not an integral basis because monodromy matrices have fractional 
rational numbers in their entries. In contrast, the sets $\Pi_{\ell}$
and $\hat{\Pi}_{\ell}$ are integral bases under the monodromy
transformations.

For the $\hat{\hat{\Pi}}$ basis, the associated monodromy matrices 
$A_{\hat{\hat{\Pi}}}$, $P_{\hat{\hat{\Pi}}}$, $T_{\hat{\hat{\Pi}}}$ 
are evaluated as
\ba
&&A_{\hat{\hat{\Pi}}}=\left(
\matrix{ -7 & 1 & 0 & 0 & 0 & 0 \cr -28 & 0 & 1 & 0 & 0 & 0 \cr 
    -84 & 0 & 0 & 1 & 0 & 0 \cr -210 & 0 & 0 & 0 & 1 & 0 \cr 
    -462 & 0 & 0 & 0 & 0 & 1 \cr -925 & 6 & -15 & 20 & -15 & 6 \cr  } 
\right)\,,\,\,\,
P_{\hat{\hat{\Pi}}}=\left(
\matrix{
   1 & 0 & 0 & 0 & 0 & 0 \cr 7 & 1 & 0 & 0 & 0 & 0 \cr 28 & 0 & 1 & 0 & 0 & 
   0 \cr 84 & 0 & 0 & 1 & 0 & 0 \cr 210 & 0 & 0 & 0 & 1 & 0 \cr 462 & 0 & 0 &
   0 & 0 & 1 \cr  } 
\right)\,,\non
&&T_{\hat{\hat{\Pi}}}=\left(
\matrix{ 6 & -15 & 20 & -15 & 6 & 
    -1 \cr 1 & 0 & 0 & 0 & 0 & 0 \cr 0 & 1 & 0 & 0 & 0 & 0 \cr 0 & 0 & 1 & 
   0 & 0 & 0 \cr 0 & 0 & 0 & 1 & 0 & 0 \cr 0 & 0 & 0 & 0 & 1 & 0 \cr  } 
\right)\,.\nom
\ea
We can study the other monodromy transformations around points
$\psi =e^{2\pi i\ell/7}$
($\ell =1,2,\cdots ,6$) by combining the results for the $\psi =1$ and 
those for $\psi =0$.
One can see that the monodromy matrices $A_{\Pi}$, $T_{\Pi}$ and 
$P_{\Pi}$ in the canonical basis have integral entries. That is a
necessary condition for the $\Pi$ to be a canonical integral basis 
because the set of the associated D-brane (integral) charge vector $Q$ is 
transformed by these monodromy transformations into 
vectors $Q\cdot A_{\Pi}$, $Q\cdot T_{\Pi}$, and $Q\cdot P_{\Pi}$. 
In this $5$-fold case, we obtain the canonical basis 
$\Pi ={\cal N}\Omega$, but the choice of the matrix $S$
has ambiguities.  They have their origins in the decomposition of the 
${\cal I}$ and the determination of the matrix $S$. 
But the data at the Gepner point
allows us to calculate the exact formula of the central charge $Z$. 
We will explain this in sections 5 and 7.

\section{Gepner Model}

There are several analyses in 3-dimensional Calabi-Yau cases 
based on the Gepner model\cite{Douglas},\cite{bCFT}. 
The boundary states, charges, and
intersection forms are discussed. Shortly we review the Gepner model and 
expand the results\cite{Douglas},\cite{bCFT} to the $d$-fold case.

A two dimensional $N=2$ minimal unitary model has a central charge 
\ba
c=\frac{3k}{k+2}\,.\nom
\ea
Here the $k$ is a positive integer and is called the level.
The primary fields of this model is labelled by a set of the 
conformal weight $h$ and the U$(1)$ charge $q$ as $(h,q)$
\ba
&&h=\frac{\ell(\ell
+2)-m^2}{4(k+2)}+\frac{s^2}{8}\,\,\,\,(\mbox{mod.}\,1)\,,\non
&&q=\frac{m}{k+2}-\frac{s}{2}\,\,\,\,(\mbox{mod.}\,2)\,.\nom
\ea
They are parametrized by a set of three integers $(\ell ,m,s)$. The 
standard range of the $(\ell ,m,s)$ is specified 
\ba
&&0\leq \ell \leq k\,,\,\,|m-s|\leq \ell\,,\non
&&s\in \{0,\pm 1\}\,\,\mbox{and}\,\,\ell +m+s\in 2{\bf Z}\,.\nom
\ea
The NS sector is associated with the $s=0$ representation while the 
$s=\pm 1$ representations belong to the Ramond sector. 
The (anti-)chiral primary states are labelled by 
$((\ell ,-\ell ,0))$ ($\ell ,+\ell ,0$) respectively in the NS sector. 
They are related to the Ramond ground states $(\ell ,\pm \ell ,\pm 1)$
by the spectral flow. 

Let us consider a Gepner model realized by tensoring $N$ minimal models
with the same level $(N-2)$.
The corresponding Landau-Ginzburg model is realized by a 
potential  
\ba
X_1^N+X_2^N+\cdots +X_N^N\,.\nom
\ea
We restrict ourselves to (complex) odd dimensional 
Calabi-Yau cases with $N=3,5,7$ ($d=1,3,5$). 
Each minimal model has a ${\bf Z}_N\times {\bf Z}_2$ symmetry whose
generators $(g,h)$s act on the primary field $\Phi^{\ell}_{m,s}$ as
\ba
&&g\Phi^{\ell}_{m,s}=\alpha^m\Phi^{\ell}_{m,s}\,\,\,\,
(\alpha =e^{2\pi i/N})\,,\non
&&h\Phi^{\ell}_{m,s}=(-1)^s\Phi^{\ell}_{m,s}\,.\nom
\ea
The ${\bf Z}_N$-symmetry is correlated with the U(1) charge. The
orbifold group of the Gepner model is generated by a diagonal ${\bf Z}_N$
generator $\prod_{j=1}^Ng_j$. Here the index ``$j$'' distinguishes the 
$N$ minimal models.
The boundary states $|\Xi\rangle\!\rangle$ which preserve $N=2$
worldsheet algebra are constructed in \cite{bCFT}. 
According to the notations of Cardy\cite{Cardy}, 
they are labelled by a set of integers 
\ba
\Xi =(L_j,M_j,S_j)\,\,\,\,\,(j=1,2,\cdots ,N)\,,\nom
\ea
for each A-, B-type boundary condition. 
The action of the discrete symmetry ${\bf Z}_N\times {\bf Z}_2$ is
expressed on the state labelled by $\Xi$ as
\ba
\left\{
\matrix{M_j &\rightarrow & M_{j}+2 & ({\bf Z}_N-\mbox{action})\cr
S_j &\rightarrow & S_{j}+2 & ({\bf Z}_2-\mbox{action})}
\right.\,.\nom
\ea
Also it is known that a physically inequivalent choice for the $S_j$ 
together with the identification under the ${\bf Z}_2$-action 
is given as 
\ba
S=\sum^{N}_{j=1}S_j\equiv 0\,\,\,\,\mbox{mod.}\,4\,.\nom
\ea
That is to say, it is enough to consider only boundary states with 
$S=0$. Also for B-type boundary states, it is known that the physically
inequivalent choices of the $M_j$ ($j=1,2,\cdots ,N$) can be described
by a sum of the $M_j$s
\ba
M=\sum^N_{j=1}M_j\,.\nom
\ea
It means that the B-type boundary states with fixed 
$L=(L_1,L_2,\cdots ,L_N)=:\{L_j\}$ are described by the single integer $M$. 
We calculate an intersection matrix $I_B$ for an $L=
(0,0,\cdots ,0)=:\{0\}$ state 
in the B-type boundary condition with $g^N=1$
\ba
I_B=(1-g^{-1})^N=
\sum^N_{\ell =0}\left(\matrix{N\cr
\ell}\right)(-1)^{\ell}g^{-\ell}\,.\nom
\ea
Then an intersection form $I_G$ in the Gepner basis is related to the $I_B$
for general $N=3,4,5,\cdots $
\ba
I_B=(1-g)(1-g^{-1})I_G\,,\,\,\,I_G=(-g^{-1})(1-g^{-1})^{N-2}\,.\nom
\ea
The $I_G$ coincides with an intersection matrix 
$m^{-1}\cdot\Sigma\cdot (m^{-1})^t$ 
in the sigma model when 
we choose the arrangement of the $\vpi_j$s appropriately
\ba
&&\vpi={}^t
(\matrix{\vpi_{a+1} &\vpi_{a+2} &\cdots &\vpi_{N-1} 
&\vpi_{0} &\vpi_{1}& \vpi_{2} &\cdots  &\vpi_{a-1}})\,.\nom
\ea
Let us recall that the basis $\Pi$ in the large volume is related to the 
$\vpi$ through the $m$
\ba
\Pi =m\vpi\,.\nom
\ea
The charge vector $Q_G$ is related to the large volume charge vector
$Q_L$ \footnote{In this section and section \ref{radius}, 
we put the super-, subscript ``$L$'' to the charge $Q$ 
in the large volume case to distinguish it from 
the $Q_G$ in the Gepner basis.} as
\ba
&&Z=Q_L\cdot \Pi =Q_G\cdot \vpi\,,\non
&&Q_L=Q_G\cdot m^{-1}\,.\nom
\ea
At a point $\psi =1$ in the moduli space of the $W$, a $d$-cycle in the
mirror shrinks into a point. An associated cycle in the M is a pure
$2d$-cycle around which a D${2d}$-brane wraps. 
The associated 
charge is specified by a charge vector $Q_L$ in the large volume
limit
\ba
Q_L=(\matrix{Q_{2d} & Q_{2d-2} & \cdots &Q_2 & Q_0})=
(\matrix{1 & 0 &0 &\cdots &0 &0})=:Q^{(0)}_L\,.\nom
\ea
When we use a Gepner basis for $\vpi$ as
\ba
\vpi={}^t(\matrix{\vpi_0 & \vpi_1 & \vpi_2 &\cdots & \vpi_{a-1} &
\vpi_{a+1} &\cdots & \vpi_{N-1}})\,,\nom
\ea
with $a=2,3,4$ for respectively $N=3,5,7$ cases, 
we can obtain a charge vector $Q^{(0)}_G$ in the Gepner basis
corresponding to the $Q^{(0)}_L$ 
\ba
Q^{(0)}_G=\left(\matrix{1&-1&0&0&\cdots &0&0}\right)\,.\nom
\ea

When we take a charge $Q_G$ of a B-type boundary state
with $|\{0\};0;0\rangle$ to be the $Q^{(0)}_G$, this boundary state is 
identified with a pure D${2d}$-brane with the charge 
$Q_L^{(0)}$. By acting a ${\bf Z}_N$-monodromy matrix $A^{-1}$, we can
change the value $M$ into an $M+2$ and obtain a charge for the state 
$|\{0\};M;0\rangle$. Also the ${\bf Z}_2$-action $h$ is implemented by
reversing a sign of the charge $Q\rightarrow -Q$. The other charges
$Q_G$ of states with $L=(\matrix{L_1 & L_2 &\cdots &L_N})$ 
can be obtained by acting 
 generators $g_j$s of the ${\bf Z}_N$-symmetries on the $Q^{(0)}_G$
\ba
Q_G=Q_G^{(0)}\prod^N_{j=1}\left(\sum^{L_j/2}_{\ell_j =-L_j/2}
g^{\ell_j}\right)\,.\label{zn}
\ea
In our case, the $N$ is odd and there is a useful relation in
calculating the $Q_G$ from the set of numbers $\{L_j\}$
\ba
g^{-1/2}=-g^{\frac{N-1}{2}}\,.\nom
\ea
Collecting all the relations, we calculate the charges for the $5$-fold
case. We list the result for the $N=7$ ($d=5$) case in the table
\ref{5dim}. 
\begin{table}[htbp]
\begin{tabular}{|c|c|c|}\hline
$L$ & $Q_G$ & $Q_L$ \\\hline
$(\matrix{0&0&0&0&0&0&0})$ & 
$(\matrix{1&-1&0&0&0&0})$ & 
$(\matrix{1&0&0&0&0&0})$\\\hline
$(\matrix{1&0&0&0&0&0&0})$ & 
$(\matrix{0&0&0&-1&1&0})$ & 
$(\matrix{5&0&2&0&-7&0})$\\\hline
$(\matrix{1&1&0&0&0&0&0})$ & 
$(\matrix{1&-1&-1&0&0&1})$ & 
$(\matrix{-3&0&-1&0&0&0})$\\\hline
$(\matrix{1&1&1&0&0&0&0})$ & 
$(\matrix{0&0&-1&-2&2&1})$ & 
$(\matrix{6&0&3&0&-14&0})$\\\hline
$(\matrix{1&1&1&1&0&0&0})$ & 
$(\matrix{2&-2&-3&-1&1&3})$ & 
$(\matrix{-5&0&-1&0&-7&0})$\\\hline
$(\matrix{1&1&1&1&1&0&0})$ & 
$(\matrix{1&-1&-4&-5&5&4})$ & 
$(\matrix{10&0&6&0&-35&0})$\\\hline
$(\matrix{1&1&1&1&1&1&0})$ & 
$(\matrix{5&-5&-9&-5&5&-9})$ & 
$(\matrix{-6&0&1&0&-35&0})$\\\hline
$(\matrix{1&1&1&1&1&1&1})$ & 
$(\matrix{5&-5&-14&-14&14&14})$ & 
$(\matrix{19&0&14&0&-98&0})$\\\hline
\end{tabular}
\caption{D-brane charges with the B-type boundary conditions.
When the set of numbers $\{L_j\}$ 
is specified, charge vectors $Q_G$ and $Q_L$ are 
determined.} 
\label{5dim} 
\end{table}

Here we use a basis $\vpi$
\ba
\vpi ={}^t(\matrix{\vpi_0 &\vpi_1 &\vpi_2 &\vpi_3 &\vpi_5 &\vpi_6  })\,.\nom
\ea
for measuring the charges in the $Q_G$ and write down results for 
the $L_j=0,1 $ ($j=1,2,\cdots ,7 $) cases for simplicity.
We make a remark here: We used the $m$ 
in Eq.(\ref{m}). But there is some arbitrariness in
determining the $m$ and the vector $Q_L$ cannot be fixed uniquely. 
Probably the arbitrariness is reduced to some 
symplectic or SO transformation and they might possibly lead to equivalent 
physics. We do not know any clear explanations about this yet.

In contrast, the result here 
shows that we can calculate a charge
vector $Q_G$ of the D-brane in the Gepner basis associated with a 
boundary state $|\{L\};M;S\rangle$
\ba
L\rightarrow Q_G\,,\nom
\ea
when a set of numbers $L=(\matrix{L_1 & L_2 &
\cdots &L_N})$ is given as an input datum.
So we can investigate relations between the $L$ and $Q_G$ more precisely. 
First we restrict ourselves to the states $|\{L\};M=0;S=0\rangle$ for
simplicity. 
Then we obtain the charge vector $Q_G$ for $N=3,4,5$ ($d=1,3,5$) cases
with respectively $a=2,3,4$ 
\ba
&&\vpi=(\matrix{\vpi_0 &\vpi_1 &\vpi_2 &\cdots &\vpi_{a-1} &\vpi_{a+1}
&\cdots &\vpi_{N-1}  })\,,\non
&&Q_G:=(\matrix{Q^{G}_{0} &Q^{G}_{1} &Q^{G}_{2}
&\cdots  &Q^{G}_{a-1} &Q^{G}_{a+1}& \cdots &Q^{G}_{N-1} } )\,,\non
&&Q^{G}_{j}=\frac{1}{N}\cdot \sum^{N-1}_{k=1}
\alpha^{-kj}(\alpha^k-1)^{-N+1}
\times \alpha^{ -\frac{k}{2}\sum^N_{j'=1}L_{j'}}\times 
\prod^N_{j''=1}\left\{\alpha^{k(L_{j''} +1)}-1\right\}\,,\,\,\,\non
&&\hspace{7cm}(j=0,1,2,\cdots ,a-1,a+1,\cdots ,N-1)\,.\nom
\ea
It is this formula that connects the set $\{L_j\}$ and the charge vector $Q_G$.
Let us recall that the D-brane charge vector $Q_L$ in the large radius 
volume with the B-type boundary condition is constructed from the $Q_G$
by the transformation matrix $m$ as $Q_L=Q_G\cdot m^{-1}$. 
The associated central charge (the BPS mass formula) $Z$ is expressed by 
the $\Pi_{\ell}$ or $\vpi$ in the appropriate parameter 
regions of the moduli space
\ba
&&Z=\sum^d_{\ell =0}Q^L_{2\ell}\cdot \Pi_{\ell}=Q_L\cdot \Pi\non
&&\qquad =(Q_Lm)\cdot (m^{-1}\Pi)\non
&&\qquad =Q_G\cdot \vpi = \wa Q^G_j\vpi_j\,.\nom
\ea
When the $\psi$ is not large (for $|\psi|<1$), 
the formula of the $Z$ is expressed in the $\vpi_j$ or $\tpi_k$ basis
\ba
&&Z=\frac{(-1)^{d+1}}{N}
\cdot \sum_{j=0,1,\cdots ,N-1}\vpi_j\cdot\sum_{k=1}^{N-1}
\alpha^{-k(j+\frac{1}{2}\sum^N_{j'=1}L_{j'})}\times
(\alpha^k-1)^{-N+1}\times\prod^N_{j''=1}
\left\{\alpha^{k(L_{j''} +1)}-1\right\}\non
&&\qquad =\frac{(-1)^N\cdot (2i)^N}{(2\pi i)^{N-1}}\times 
\sum^{N-1}_{k=1}\tpi_k \cdot (-1)^k \times \prod^N_{j=1}
\left\{\sin \frac{\pi k(L_j +1)}{N}\right\}\,.\label{gcenter}
\ea

Next the state $|\{L\};M=2\ell ;0\rangle$ $(\ell \neq 0)$ 
is obtained by acting the 
${\bf Z}_N$ monodromy matrix $A^{-1}$ 
on the Z in the $|\{L\};M=0;0\rangle$ case
\ba
Z=Q\cdot (A^{-1}_{\Pi})^{\ell}\Pi\,.\nom
\ea
The $S\neq 0$ case can be realized by acting 
the remaining ${\bf Z}_2$ symmetry on 
the charge $Q\rightarrow -Q$.

\section{Cycles}

In this section, we study cycles associated with the periods $\varpi_j$
and their intersection forms.
The cycles are susy $d$ cycles in the $d$ fold consistent with the A-type 
boundary condition.

\subsection{Cycles}
For the $d$-fold variety $W$
\ba
&&W\,;\,\widehat{\{p=0\}/{\bf Z}_N^{\otimes (N-1)}}\,,\non
&&p=X^N_1+X^N_2+\cdots +X^N_N-N\psi X_1X_2\cdots X_N=0\,,\nom
\ea
we consider a set of $d$ chains
$V_j(\psi)$ $(j=0,1,2,\cdots ,N-1)$
\ba
&&V_j(\psi)=\left\{
\begin{array}{l}
X_1,X_2,\cdots ,X_{N-1}\,;\,\mbox{real positive}\\
X_N=1\,,\\
\dis X_{N-1}\,\,\mbox{s.t.}\,\,\,\mbox{arg}\,X_{N-1}\rightarrow 
\pi \cdot \frac{2j+1}{N}\,\,\,\mbox{as}\,\,\psi\rightarrow 0\,\,\,
\end{array}\right\}
\,.\nom
\ea
In this definition, the index ``$j$'' of the $V_j$ is defined modulo 
$N$, that is, $V_{N+j}=V_j$ for an arbitrary $j\in {\bf Z}$.
These are chains with the same boundary under an identification of the 
discrete symmetry ${\bf Z}_N^{\otimes (N-1)}$. But an arbitrary combination 
$V_i-V_j$ ($i,j=0,1,2,\cdots ,N-1$) is a $d$-cycle.
These cycles are related with the $V_k$s through
an intersection matrix $I_{j,k}$
\ba
&&{\cal C}_j=\sum^{N-1}_{k=0}I_{j,k}V_k\,,\,\,\,
I_{j,k}=\frac{1}{N}\sum^{N-1}_{\ell =0}\alpha^{\ell (j-k)}
(1-\alpha^{\ell})^{N-2}\,.\label{cycle}
\ea
More explicitly, this expression is evaluated as a linear combination of
$d$-cycles 
\ba
&&{\cal C}_j=\sum^{N-2}_{n=0}(-1)^n\left(\matrix{N-2\cr
n}\right)V_{j+2}\,\,\,\,\,
(j\,;\,\mbox{mod.}\,N)\,\non
&&\qquad =\sum^{N-3}_{n=0}(-1)^n\left(\matrix{N-3\cr n}\right)
(V_{j+n}-V_{j+n+1})
\,\,\,\,\,
(j\,;\,\mbox{mod.}\,N)\,.\nom
\ea
Let us consider the $N=$odd cases for $\psi =0$.
In the $N=$odd case, a phase of $X_{N-1}$ 
in the $V_j$ tends to a definite value
for $\psi \rightarrow 0$
\ba
&&
\mbox{arg}\,X_{N-1}\rightarrow 
\pi +\frac{2\pi}{N}\left(j-\frac{N-1}{2}\right)\qquad (N=\mbox{odd})\,.\nom
\ea
The $d$ chain $V_j$ for $N=$odd is 
represented in this limit as
\ba
&&\mbox{Im}\,X_{\ell}=0\qquad (\ell =1,2,\cdots ,N-2,N)\,,\non
&&\mbox{Im}\,\left(\alpha^{\frac{N-1}{2}-j}X_{N-1}\right)=0\,.\non
&&\frac{N-1}{2}-j\in {\bf Z}\,,\nom
\ea
At this special moduli point, there are several analyses about 
the cycles in \cite{becker,hori}. 
The associated susy $d$ chain for the orbifold point is 
obtained in \cite{becker} as
\ba
\mbox{Im}\,\left(\alpha^{m_{\ell}}X_{\ell}\right)=0\,,\,\,\,
m_{\ell}\in {\bf Z}\qquad 
(\ell =1,2,\cdots, N)\,.\label{beck}
\ea
An arbitrary susy cycle is constructed as a difference of 
any two susy chains (\ref{beck}) at this moduli point. 
In our case, the  $V_j$ evaluated at $\psi =0$ 
corresponds to a special kind of susy chains
labelled by  a set of numbers $\{m_{\ell}\}$ at the orbifold point 
\ba
m_{\ell}=\left(\frac{N-1}{2}-j\right)\delta_{\ell ,N-1}\,.\nom
\ea
But we obtain a formula of susy cycles 
in the $d$-fold $W$ 
at a specified moduli point $\psi$.

\subsection{Periods}

Next we introduce 
a set of periods $q_j(\psi)$ associated with the ${\cal C}_j$.
These $q_j(\psi)$s are related with our periods $\tpi_k(\psi)$ as
\ba
q_j(\psi)=\frac{1}{2\pi i\cdot N^{N-1}}\cdot 
\sum^{N-1}_{k=1}
\alpha^{jk}(\alpha^k -1)^{N-1}\tpi_k(\psi)\,\,\,\,\,\,(j=0,1,2,\cdots
,N-1)\,.\label{findc}
\ea
In order to prove the Eq.(\ref{findc}), we introduce a set of one-cycles 
$\gamma_j$ as unions of half-lines
\ba
\gamma_j=\{\mbox{arg}\,X_j=0\}\cup
\{\mbox{arg}\,X_j=\frac{2\pi}{N}\}\,\,\,\,
(j=1,2,\cdots ,N-2)\,.\nom
\ea
From these $d$ one-cycles, we can construct $d$-cycles ${\cal C}_j$
\ba
&&{\cal C}_j=\{ (X_1,X_2,\cdots ,X_{N-2},X_{N-1},1)\,;\,\non
&&\qquad \qquad (X_1,X_2,\cdots ,X_{N-2})\in \gamma_1\times
\gamma_2\times\cdots \times\gamma_{N-2}\,,\non
&&\qquad\qquad \mbox{branch of}\,\, X_{N-1}\,\,
\mbox{is specified by ``$j$''}\}
\,\,\,\,\,(j=0,1,2,\cdots ,N-2)\,.\nom
\ea
We calculate a period $q_0(\psi)$ associated with the ${\cal C}_0$
\ba
&&q_0(\psi)=\sum^{\infty}_{m=1}u_m(N\psi)^m{\bf I}_m\,,\non
&&u_m=-\frac{1}{N}\cdot \alpha^{m/2}\cdot 
\frac{\dis \Gam\!\left(\frac{N-1}{N}\cdot m\right)}
{\dis \Gam\!\left(1-\frac{m}{N}\right)\Gam\!\left(m\right)}\,,\non
&&{\bf I}_m=\int_{\gamma_1\times\gamma_2\times  \cdots \times \gamma_{N-2}}
dX_1dX_2\cdots dX_{N-2}\frac{(X_1X_2\cdots X_{N-2})^{m-1}}
{\Delta^{\frac{N-1}{N}m}}\,,\non
&&\Delta :=1+X^N_1+X^N_2+\cdots +X^N_{N-2}\,.\nom
\ea
This ${\bf I}_m$ is transformed into an integral associated with the 
$V_0$
\ba
{\bf I}_m=(1-\alpha^m)^{N-2}\cdot 
\int^{\infty}_{0}dX_1\int^{\infty}_{0}dX_2\cdots
\int^{\infty}_{0}dX_{N-2}
\frac{(X_1X_2\cdots X_{N-2})^{m-1}}
{\Delta^{\frac{N-1}{N}m}}\,.\nom
\ea
When we recall the action of the ${\bf Z}_N$ monodromy transformation
${\cal A}$ around the $\psi =0$
\ba
{\cal A}\,;\,(N\psi)^m\rightarrow \alpha^m(N\psi)^m\,,\nom
\ea
the $q_0(\psi)$ is expressed as
\ba
&&q_0(\psi)=\sum^{\infty}_{m=1}u_m(N\psi)^m
\int_{\gamma_1\times\gamma_2\times  \cdots \times \gamma_{N-2}}
dX_1dX_2\cdots dX_{N-2}\frac{(X_1X_2\cdots X_{N-2})^{m-1}}
{\Delta^{\frac{N-1}{N}m}}\non
&&\qquad =(1-{\cal A})^{N-2}\cdot
\sum^{\infty}_{m=1}u_m(N\psi)^m
\int^{\infty}_{0}dX_1\int^{\infty}_{0}dX_2\cdots
\int^{\infty}_{0}dX_{N-2}
\frac{(X_1X_2\cdots X_{N-2})^{m-1}}
{\Delta^{\frac{N-1}{N}m}}\,.\nom
\ea
From this formula, 
we obtain a relation between the ${\cal C}_0$ and the $V_0$
\ba
{\cal C}_0=(1-{\cal A})^{N-2}V_0\,.\nom
\ea
On the other hand, the $q_0(\psi)$ can be expressed in a series
expansion explicitly
\ba
&&q_0(\psi)=\left(\frac{2\pi i}{N}\right)^{N-2}\cdot \frac{1}{N}\cdot
(-1)^{N-1}\non
&&\qquad \times \sum^{\infty}_{m=1}
\frac{\dis \Gam \left(\dis \frac{m}{N}\right)}
{\Gam (m)\left[\Gam \left(\dis 1-\frac{m}{N}\right)\right]^{N-1}}
\times (\alpha^{\frac{N-1}{2}}\cdot N\psi)^m\,.\nom
\ea
It is related with a period $\vpi_0(\psi)$ as
\ba
&&q_0(\psi)=(-1)^N\cdot \left(\frac{2\pi
    i}{N}\right)^{N-2}\varpi_0(\psi)\,,\non
&&\varpi_0(\psi)=
-\frac{1}{N}\cdot
\sum^{\infty}_{m=1}
\frac{\dis \Gam \left(\dis \frac{m}{N}\right)}
{\Gam (m)\left[\Gam \left(\dis 1-\frac{m}{N}\right)\right]^{N-1}}
\times (\alpha^{\frac{N-1}{2}}\cdot N\psi)^m\,.\nom
\ea
Also a relation $q_j(\psi)=q_0(\alpha^j\psi)$ means that each 
cycle ${\cal C}_j$ is related with a period $q_j$.
That completes the proof of the statement in Eq.(\ref{findc}).

We make a comment on the branch of the $X_{N-1}$: First we
introduce a variable $v=N\psi \Delta^{-\frac{N-1}{N}}X_1X_2\cdots X_{N-2}$.
In the small $v$ case ($v=0$), the $X_{N-1}$ is evaluated as
\ba
X_{N-1}=\Delta^{1/N}\cdot \exp\left(\frac{\pi i}{N}(2j'+1)\right)
\,\,\,\,
(j'=0,1,2,\cdots ,N-1)\,.\nom
\ea
The argument of $X_{N-1}$ corresponds to the branch of the chain $V_{j'}$.
We find an exact representation of the $X_{N-1}$
\ba
&&\Delta^{-1/N}\cdot X_{N-1}=-\sum^{\infty}_{n=0}
\frac{\alpha^{(n+1)\left(j'+\frac{1}{2}\right) }}{N}\cdot 
\frac{\dis \Gam \left(\frac{N-1}{N}(n+1)-1\right)}{\dis \Gam (n+1)\Gam 
  \left(1-\frac{n+1}{N}\right)}\cdot v^n\,,\non
&&\hspace{6.7cm}(j'=0,1,2,\cdots ,N-1)\,.\nom
\ea
On the other side, the $X_{N-1}$ is evaluated in the large $v$ region
\ba
X_{N-1}=\Delta^{1/N}\cdot 
v^{\frac{1}{N-1}}\exp\left(\frac{2\pi
    i}{N-1}{j''}\right)\,\,\,\,
({j''}=0,1,2,\cdots ,N-1)\,.\nom
\ea
When we move the parameter $\psi$ from $0$ to $\infty$ on the real line, 
two of the ($N-1$) $X_{N-1}$s with $j=0$ and $N-1$ 
coincide at $v=N(1-N)^{\frac{1}{N}-1}$.
Its value is evaluated as $X_{N-1}=\Delta^{1/N}\cdot (N-1)^{-1/N}$. 
The other $X_{N-1}$s do not collide one another for the $\psi\in {\bf R}_{+}$.

Next we will consider cycles associated with the $\tpi_{\ell}$. 
Because these $q_j(\psi)$ are related with our periods $\tpi_k(\psi)$ 
in Eq.(\ref{findc}), 
we can obtain cycles $\tilde{C}_{\ell}$ 
associated with the periods $\tpi_{\ell}$
as
\ba
&&\tilde{C}_{\ell}
=\frac{1}{(1-\alpha^{\ell})^{N-1}}\sum^{N-1}_{j=0}\alpha^{-j\ell}
{\cal C}_j\non
&&\qquad =
\frac{1}{1-\alpha^{\ell}}\sum^{N-1}_{k=0}\alpha^{-\ell k}V_k 
\,\,\,\,\,\,(1\leq \ell \leq N-1)\,.\nom
\ea
These cycles $\tilde{C}_{\ell}$s 
diagonalize the action of the ${\bf Z}_N$ monodromy 
$\psi\rightarrow \alpha\psi$
\ba
\tilde{C}_{\ell}\rightarrow \alpha^{\ell}\tilde{C}_{\ell}\,,\nom
\ea
but the sets of ${\cal C}_j$ and $V_k$ change cyclically
\ba
Q_j\rightarrow Q_{j+1}\,,\,\,\,
V_k\rightarrow V_{k+1}\,.\nom
\ea
The $\tilde{C}_{\ell}$s belong to H${}_d(W\,;\,{\bf C})$. On the 
other hand, the ${\cal C}_j$s and $V_k$s are rational 
homology cycles in H${}_d(W\,;\,{\bf Q})$.

The results of the CFT at the Gepner point imply that an 
intersection matrix associated with the cycles ${\cal C}_j$ must be 
the $I_B$
\ba
&&{\cal C}_j\cap {\cal C}_{j'}=(I_B)_{j,j'}
=\frac{1}{N}\cdot
\sum^{N-1}_{r=0}\alpha^{r(j+j'+1)}(1-\alpha^r)^{N-2}\,,
\,\,\,\,\,(j,j'=0,1,2,\cdots ,N-1)\,.\nom
\ea
This fact allows us to calculate intersections of the 
$V_k$s formally by using the Eq.(\ref{cycle})
\ba
V_k\cap V_{k'}=\frac{(-1)^N}{N}\cdot \sum^{N-1}_{r=1}
\frac{\alpha^{r(-k+k'-1)}}{(1-\alpha^r)^{N-2}}\,,\,\,\,\,\,\,
(k,k'=0,1,2,\cdots ,N-1)\,.\nom
\ea
By using this relation, we can show that 
the $d$-cycles $V_{k,\ell}:=V_k-V_{\ell}$s have intersection forms
\ba
&&V_{k\ell}\cap V_{k'\ell'}=
\frac{(-1)^N}{N}\cdot \sum^{N-1}_{r=1}
\frac{\alpha^{-r}}{(1-\alpha^r)^{N-2}}
\times (\alpha^{-kr}-\alpha^{-\ell r})(\alpha^{k'r}-\alpha^{\ell' r})
\,,\non
&&\hspace{6.5cm}
(k,\ell ,k' ,\ell'=0,1,2,\cdots ,N-1)\,.\nom
\ea

\section{Central Charge in Large Radius Region}\label{radius}

In this section, we calculate the B-type central charge 
$Z$ for Calabi-Yau $d$-fold from the point of view of mirror symmetry. 
We analyze the structure of the $Z$ at a generic {\kae }
structure moduli point near the large radius region. 
Also we consider relations with this formula with 
the results in the Gepner model.

\subsection{B-type Central Charge }

In this subsection, we consider the B-type 
central charge $Z$ in the large radius 
region of the $M$.  
Let us recall that it is a product of a charge vector $Q_L$ and
canonical basis $\Pi$
\ba
&&Z=Q_L\cdot \Pi=\tilde{Q}\cdot \Omega\,,\non
&&\tilde{Q}=Q_L{\cal N}\,.\label{z}
\ea
The matrix ${\cal N}$ transforms the charges $Q_L$ into $\tilde{Q}$ by a
fractional redefinition of the charge lattices.
When we consider the $d$-fold $M$, the D${}_{2d-2p}$ brane charge is 
given by integrating the $F^p$ (more precisely, $2p$ form parts of
an associated Mukai vector $v({\cal E})$ of a sheaf ${\cal E}$)
over 
the $\Sigma_{2p}$ with the B-type boundary
conditions ($p=0,1,2,\cdots ,d$)
\ba
&&Q_{2d-2p}=\int_{\Sigma_{2p}}e^F\,.\nom
\ea
We prepare several notations and concrete formulae in order to write the
central charge $Z$ explicitly.
First the function $\vpi\left(+\frac{\rho}{2\pi i}\right)$ is defined 
by using a formal parameter $\rho$ with $\rho^{N-1}=0$
\ba
&&\vpi_0^{-1}\times \vpi\left(\frac{\rho}{2\pi i}\right):=e^{t\rho}\times 
\exp\left(\sum_{n\geq 2}\rho^n x_n\right)\,,\non
&&2\pi i t
=\log z+\frac{\dis\sum^{\infty}_{n=1}\frac{(Nn)!}{(n!)^N}
\left(\sum^{Nn}_{m=n+1}\frac{N}{m}\right)z^n}
{\dis \sum^{\infty}_{n=0}\frac{(Nn)!}{(n!)^N}z^n}\,,\,\,\,
q=e^{2\pi i t}\,,\non
&&x_n=\frac{1}{(2\pi i)^n}\frac{1}{n!}\del^n_{\rho }\log
\left[\sum^{\infty}_{m=0}\frac{\Gam\!(N(m+\rho)+1)}
{\Gam\!(N\rho +1)}\left(\frac{\Gam\!(\rho +1)}
{\Gam\!(m+\rho +1)}\right)^N z^m\right]\Biggr|_{\rho =0}\,.\nom
\ea
Also the $\hat{K}(\rho)$ is related to Riemann's zeta functions 
\ba
&&\hat{K}(\rho)
=\exp\left[2\sum_{m=1}\frac{N-N^{2m+1}}{2m+1}\zeta (2m+1)
\left(\frac{\rho}{2\pi i}\right)^{2m+1}
\right]\,.\nom
\ea
Then the $\hat{\Pi}_m$ and $\hat{\hat{\Pi}}_n$ is expressed as
$(0\leq m\leq N-2\,;\,0\leq n\leq N-2)$ 
\ba
&&\vpi_0^{-1}\times 
\sqrt{\hat{K}(-\rho)}\cdot \vpi\left(-\frac{\rho}{2\pi i}\right)=
e^{-\rho t}\times \exp \left(\sum_{n\geq 2}(-\rho)^n \cdot
y_n\right)\,,\non
&&\hat{k}_{2m+1}=\frac{1}{(2\pi i)^{2m+1}}\cdot 
\frac{N-N^{2m+1}}{2m+1}\zeta (2m+1)\,,\non
&&y_{2m}:=x_{2m}\,,\,\,\,y_{2m+1}:=x_{2m+1}+\hat{k}_{2m+1}\,,\,\,\,
(m=1,2,\cdots)\,,\non
&&\vpi_0^{-1}\times \hat{\Pi}_m =
\left[\sqrt{\hat{A}(+\rho)}\cdot (e^{\rho}-1)^m
\cdot 
e^{-\rho t}\times \exp \left(\sum_{n\geq 2}(-\rho)^n \cdot
y_n\right)
\right]_{\rho^{N-2}}\,,\label{pi1}\\
&&\vpi_0^{-1}\times 
\hat{\hat{\Pi}}_n=\left[\sqrt{\hat{A}(+\rho)}\cdot e^{n\rho}
\cdot 
e^{-\rho t}\times \exp \left(\sum_{n\geq 2}(-\rho)^n \cdot
y_n\right)
\right]_{\rho^{N-2}}\,.\label{pi2}
\ea

Now we will rewrite the formula Eq.(\ref{z}) in the bases $\hat{\Pi}$ and 
$\hat{\hat{\Pi}}$. 
The $\hat{\Pi}$ and $\hat{\hat{\Pi}}$ are 
related to the $\Pi_n$ and $\Omega_{\ell}$ $(0\leq n\leq N-2\,;\,0\leq
\ell \leq N-2)$
\ba
&&\Pi_n =\sum_{m=0}^{N-2}S_{n,m}\hat{\Pi}_m\,,\,\,\,
\hat{\Pi}_n=\sum_{m=0}^{N-2}U_{n,m}\hat{\hat{\Pi}}_m\,,\,\,
(0\leq n\leq N-2)\,,\non
&&\hat{\Pi}_m=\sum_{\ell =0}^{N-2}V_{m,\ell}\Omega_{\ell}\non
&&U_{n,m}=\left(\matrix{n\cr m}\right)\cdot (-1)^{n-m}\,,\,\,\,
(0\leq n\leq N-2\,;\,0\leq m\leq N-2)\,,\non
&&V_{m,\ell}:=\left[\sqrt{\hat{A}(+\rho)}\cdot (e^{\rho}-1)^m
\cdot (-\rho)^{\ell}\right]_{\rho^{N-2}}\,,\,\,\,
(0\leq m\leq N-2\,;\,0\leq \ell\leq N-2)\,.\nom
\ea
We can obtain the central charge by using these equations 
Eqs.(\ref{pi1}),(\ref{pi2})
\ba
&&Z=\sum_{\ell =0}^{N-2}
Q^L_{2\ell}\cdot
\left[\sqrt{\hat{A}(+\rho)}\cdot\left\{
\sum_{m=0}^{N-2}S_{\ell ,m}(e^{\rho}-1)^m\right\}
\cdot  
e^{-\rho t}\times \exp \left(\sum_{n\geq 2}(-\rho)^n \cdot
y_n\right)\right]_{\rho^{N-2}}\,\non
&&\!\qquad =\sum_{\ell =0}^{N-2}
Q^L_{2\ell}\cdot
\left[\sqrt{\hat{A}(+\rho)}\cdot\left\{
\sum_{m,n=0}^{N-2}S_{\ell ,m}U_{m,n} \right\}\cdot e^{n\rho}
\cdot  
e^{-\rho t}\times \exp \left(\sum_{n\geq 2}(-\rho)^n \cdot
y_n\right)\right]_{\rho^{N-2}}\,.\label{center}
\ea
These formulae Eqs.(\ref{center}) remind us the couplings of RR fields
(potentials) $C$ to 
the gauge fields on the D-brane, that is to say, a Chern-Simons term
in the B-model
\ba
\int C\wedge e^{\cal F}\wedge \sqrt{\frac{\hat{A}(M)}{\hat{A}(N)}}\,.\nom
\ea
Here the $N$ is a normal bundle of the world volume. 
The $C$ couples with the Mukai vector
\ba
e^{\cal F}\wedge \sqrt{\frac{\hat{A}(M)}{\hat{A}(N)}}\,.\nom
\ea
and the gauge field $F$ is combined with the 
Kalb-Ramond field $B$ into the ${\cal F}$
\ba
{\cal F}=F-B\,.\nom
\ea
In the context of the topological sigma model, 
this $B$-field is combined into a complexified {\kae} form
\ba
&&B+iJ=t\,[D]\,\,\,([D]\in\mbox{H}^2(M))\,.\non
&&B=\mbox{Re}(t)[D]\,,\,\,\,J=\mbox{Im}(t)[D]\,.\nom
\ea
In our case, the ``$[D]$'' is a $1$st Chern class of a hyperplane bundle 
of $CP^{N-1}$
\ba
[D]=c_1\left({\cal O}_{CP^{N-1}}(1)\right)\,,\label{sheaf}
\ea
and then the ${\cal F}$ is represented as
\ba
&&{\cal F}=F-B=F-\mbox{Re}(t)[D]\,.\nom
\ea
When one shifts the B-field, the real part of the $t$ changes, for an example, 
$t\rightarrow t+1$
and the ${\cal F}$ changes.
It implies a coupling
\ba
C\wedge e^{-Re(t)\,[D]}\wedge \sqrt{\frac{\hat{A}(M)}{\hat{A}(N)}}\,.\nom
\ea
In our case, we see an analogous term in the $Z$ in Eqs.(\ref{center})
\ba
&&Z=\sum_{\ell =0}^{N-2}Q^L_{2\ell}\cdot
\left[\sqrt{\hat{A}(\rho )}
\cdots e^{-\rho t }\times \cdots\right]_{\rho^{N-2}}\,.\nom
\ea
Up to a multiplicative constant, 
this formal parameter $\rho$ could be identified with the divisor $[D]$
and the $e^{-t\rho}$ is represented by using the  $[D]$
\ba
e^{- t\rho }=e^{-t\,[D]}=e^{-(iJ+B)}\,.\nom
\ea
Then the $Z$ is expressed as integral formulae 
\ba
&&Z=\sum_{\ell =0}^{N-2}Q^L_{2\ell}\cdot
\int_M \Biggl[\sqrt{\hat{A}([D])}\cdot\left\{
\sum_{m=0}^{N-2}S_{\ell ,m}(e^{[D]}-1)^m\right\}
\cdot  
e^{- t[D]}\non
&&\qquad \qquad 
\times \exp \left(\sum_{n\geq 2}(-[D])^n \cdot
y_n\right)\Biggr]
\times \left(\int_M[D]^d\right)^{-1}\,\label{div1}\\
&&\qquad =\sum_{\ell =0}^{N-2}Q^L_{2\ell}\cdot
\int_M\Biggl[\sqrt{\hat{A}([D])}\cdot\left\{
\sum_{m,n=0}^{N-2}S_{\ell ,m}U_{m,n} \right\}\cdot e^{n[D]}
\cdot  
e^{-t[D]}\non
&&\qquad \qquad  
\times \exp \left(\sum_{n\geq 2}(-[D])^n \cdot y_n\right)\Biggr]
\times \left(\int_M[D]^d\right)^{-1}\,.\label{div2}
\ea
Here the integral over the $M$ is defined with $[H]:=N[D]$
\ba
&&\int_M(\cdots) =\int_{CP^{N-1}}(\cdots)\cdot [H]\,.\nom
\ea
In the formula of the $Z$ in Eq.(\ref{div2}), 
there appears a term $e^{n[D]}$. When we 
identify the $\rho$ with the $[D]$, the $n[D]$ is identified 
with $c_1\left({\cal O}_{CP^{N-1}}(-n)^{\ast}\right)$. 
Then the term $e^{n[D]}$
in Eq.(\ref{div2})
is interpreted as a Chern character of the line bundle 
${\cal O}_{CP^{N-1}}(-n)^{\ast}$ 
\ba
e^{n[D]}=\ch \left({\cal O}_{CP^{N-1}}(-n)^{\ast}\right)\,.\nom
\ea
We will explain this geometrical interpretation about $Z$ in the 
next subsection. 

As a last topic in this subsection, we consider a monodromy 
transformation around large radius point $\mbox{Im}t =\infty$ 
(equivalently, $\psi = \infty$).
When one performs a monodromy transformation around the $\psi =\infty$, 
the $t$ is shifted into $t+1$ and the 
$B$-field changes as $B\rightarrow B+[D]$. 
In the $Z$, the $x_n$s are invariant under the integral shift of the $t$, 
but the term $e^{-t[D]}$ is transformed into $e^{-[D]}\cdot e^{-t[D]}$. 
This shift affects on the $Z$
\ba
&&Z=\sum_{\ell =0}^{N-2}Q^L_{2\ell}\cdot
\left[
\cdots \sum_{m=0}^{N-2}S_{\ell ,m}
(e^{[D]}-1)^m e^{-t[D] }\times \cdots\right]\,\non
&&\qquad \rightarrow 
\sum_{\ell =0}^{N-2}Q^L_{2\ell}\cdot
\left[
\cdots 
\sum_{m=0}^{N-2}S_{\ell ,m}
\sum_{m'=0}^{N-2}(-1)^{m'} (e^{[D]}-1)^{m+m'} e^{-t[D] }\times
\cdots\right]\,,\non
&&Z=\sum_{\ell =0}^{N-2}Q^L_{2\ell}\cdot
\left[
\cdots \left\{\sum_{m,n=0}^{N-2}S_{\ell ,m}U_{m,n}\right\}
e^{n[D]} e^{-t[D] }\times \cdots\right]\,\non
&&\qquad \rightarrow 
\sum_{\ell =0}^{N-2}Q^L_{2\ell}\cdot
\left[
\cdots \left\{\sum_{m,n=0}^{N-2}S_{\ell ,m}U_{m,n}\right\}
e^{(n-1)[D]} e^{-t[D] }\times
\cdots\right]\,.\nom
\ea
It induces rearrangements of the components of the matrices $S$ and
$U$ and leads to a reshuffle of the charge vector.
Also it gives us  monodromy matrices on the basis $\hat{\Pi}$ and 
$\hat{\hat{\Pi}}$. 
Especially monodromy matrices $T_{\hat{\Pi}} $ and $T_{\Omega}$  
around $\psi =\infty$
can be obtained in the $\hat{\Pi}$ and $\Omega$ bases 
by this consideration
\ba
&& \hat{\Pi}_m =\vpi_0 \times
\left[\sqrt{\hat{A}(\rho)}\cdot (e^{\rho}-1)^m
\cdot 
e^{-\rho t}\times \exp \left(\sum_{n\geq 2}(-\rho)^n \cdot
y_n\right)
\right]_{\rho^{N-2}}\non
&&\qquad \rightarrow \vpi_0\times 
\left[\sqrt{\hat{A}(\rho)}\cdot
\sum_{m'=0}^{N-2}(-1)^{m'} (e^{\rho}-1)^{m+m'}
\cdot 
e^{-\rho t}\times \exp \left(\sum_{n\geq 2}(-\rho)^n \cdot
y_n\right)
\right]_{\rho^{N-2}}\,,\non
&&\Omega_{\ell}=\vpi_0\times 
\left[e^{\rho t}\times \exp \left(\sum_{n\geq 2} \rho^n \cdot
y_n\right)\right]_{\rho^{\ell}}\,\non
&&\qquad \rightarrow \vpi_0\times 
\left[e^{\rho}\cdot 
e^{\rho t}\times \exp \left(\sum_{n\geq 2} \rho^n \cdot
y_n\right)\right]_{\rho^{\ell}}\,,\non
&&T_{\hat{\Pi}}=\sum^{N-2}_{\ell =0}(-{\cal T})^{\ell}
\,,\,\,\,T_{\Omega}=e^{{\cal T}^t}\,,\non
&&{\cal T}_{\ell ,\ell'}=\delta_{\ell +1,\ell'}
\,,\,\,\,\,\,\,(0\leq \ell\leq N-2\,;\,0\leq \ell'\leq N-2\,)\,.\nom
\ea
The formulae of these $T_{\hat{\Pi}}$ and $T_{\Omega}$ for $N=7$ case coincide 
with those in section 4.

\subsection{Relation with Results in the Gepner Model}

In this subsection, we investigate the $Z$ 
associated with the charge $\{Q^G_j\}$ at the Gepner point  
and study relations with
the B-type boundary state $|\{L\};M;S\rangle$
in the Gepner model. 
The $Z$ is obtained in Eq.(\ref{gcenter}) by using the set 
$\{Q^G_j\}$ $(j=0,1,2,\cdots ,N-1)$.
We can rewrite this formula as
\ba
&&Z=
N^{-1}\cdot \sum^{N-1}_{k=1}
(\alpha^k-1)^{N-1}
\left(\sum^{N-1}_{j=0}Q^G_j\alpha^{kj}\right)
\int_M\Biggl[\,\frac{\alpha^{k}}{e^{[D]}-\alpha^k}\cdot 
\sqrt{\hat{A}([D])}
\cdot e^{-t[D]}\non
&&\qquad\qquad \times 
\sqrt{\hat{K}(-[D])}
\times \exp \left(\sum_{n\geq 2}(-[D])^n \cdot
x_n\right)
\Biggr]\label{sheaf2}\\
&&\qquad =
N^{-1}\cdot 
\sum^{N-2}_{n=0}\left\{
\sum^{N-2}_{\ell =n}\sum^{N-1}_{j=0}
Q^G_jP_{j,\ell}\cdot \left(\matrix{\ell \cr n}\right)(-1)^{\ell -n}
\right\}
\int_M\Biggl[\,
e^{n[D]} \cdot 
\sqrt{\hat{A}([D])}\cdot 
e^{-t[D]}\non
&&\qquad\qquad \times 
\sqrt{\hat{K}(-[D])}
\times \exp \left(\sum_{m\geq 2}(-[D])^m \cdot
x_m\right)\Biggr]\,,\non
&&P_{j,\ell}=\delta_{\ell ,N-2}+N\cdot 
(-1)^{j-\ell}\left(\matrix{N-2-\ell\cr N-1-j}\right)\,,\,\,\,
\alpha =e^{2\pi i/N}\,,\nom
\ea
Here $\dis \sum^{N-1}_{j=0}Q^G_jP_{j,m}$s are integers because 
the charges $Q^G_j$ and the $P_{j,m}$s are integers.
Information about charges 
is encoded in a function $R$
\ba
R([D]):=N^{-1}\cdot \sum^{N-1}_{k=1}
(\alpha^k-1)^{N-1}
\left(\sum^{N-1}_{j=0}Q^G_j\alpha^{kj}\right)
\cdot \frac{\alpha^{k}}{e^{[D]}-\alpha^k}\,.\nom
\ea
Together with the $\sqrt{\hat{A}}$,
analogs of the Mukai vector appear in the $Z$ 
\ba
&&\sqrt{\hat{A}([D])}\cdot R([D])\,,\non
&&\sqrt{\hat{A}([D])}\cdot e^{n[D]} 
\,.\nom
\ea
We investigate this structure in the geometrical point of view. 
The Calabi-Yau $d$-fold $M$ is embedded in the ambient projective space 
$X=CP^{N-1}$. A line bundle ${\cal O}_X(m)$ of the $X$
is defined as an $m$ tensor product of a section of a hyperplane bundle.
Here we introduce a set of line bundles $R_a=\co_X (-a[D])$.
An intersection paring $I_{a,b}$ between these bundles $R_a$ and $R_b$
is defined as a Euler characteristic 
$\chi_X(R_a,R_b)$
\ba
I_{a,b}=\chi_X(R_a,R_b)=\int_X\ch (R_a)^{\ast}\ch (R_b)\td (TX)\,.\nom
\ea
We can evaluate the $I_{a,b}$ and its inverse explicitly
\ba
&&I_{a,b}=\chi_{X} (R_a,R_b)=
\left(\matrix{d+1+a-b\cr a-b}\right)\,,\non
&&(I^{-1})_{a,b}=(-1)^{a-b}
\left(\matrix{N\cr a-b}\right)\,,\,\,\,(N=d+2)\,.\nom
\ea
Next we define a set of dual basis $\{S_a\}$ for the bundles $\{R_a\}$
\ba
&&\ch (S_a):=\sum_b (I^{-1})_{b,a} \ch (R_b)
=\sum_b (-1)^{b-a}\left(\matrix{N\cr b-a}\right)\ch (R_b)
\,,\non
&&\ch (S_a^{\ast}):=\sum_b (I^{-1})_{a,b} \ch (R_b^{\ast})
=\sum_b (-1)^{a-b}\left(\matrix{N\cr a-b}\right)\ch (R_b^{\ast})
\,.\nom
\ea
The two sets of bases are orthonormal with respect to  
the intersection paring 
\ba
&&\langle R_a,S_b \rangle =\chi_X (R_a,S_b)=
\delta_{a,b}\,,\non
&&\langle S_a,R_b \rangle =\chi_X (S_a,R_b)=
\delta_{a,b}\,.\nom
\ea
The set of line bundles (sheaves) 
$\{\co (a)\}$ $(a=0,1,2,\cdots ,N-1)$ is an exceptional collection
of the $CP^{N-1}$ and also turns out to be a foundation of an
associated helix of $CP^{N-1}$. We can consider a left mutation 
${\bf L}$ on the
set $\{\co (a)\}$ because a condition 
$\mbox{Ext}^0(\co (a-1),\co (a))=
H^0(\co (a-1),\co (a))\neq 0$ is satisfied.
It leads to a relation of Chern characters of the bundles
\ba
\ch ({\bf L}_{a-1}\co (a))=
\ch (\co (a))-\chi (\co(a-1),\co (a))\ch (\co (a-1))\,,\label{mutation}
\ea
where we introduce an abbreviated notation ${\bf L}_{a-1} \co (a):=
{\bf L}_{\co (a-1)}\co (a)$.
By using this formula Eq.(\ref{mutation}) iteratively, 
we obtain a relation for the $S^{\ast}_a$
\ba
\ch ({\bf L}_0{\bf L}_1\cdots {\bf L}_{a-1}\co (a))
=\sum^a_{b=0}(-1)^{a-b}\left(\matrix{N\cr a-b}\right)\ch (\co (b))=
\ch (S^{\ast}_a)\,.\nom
\ea
Thus each element $S^{\ast}_a$ of the dual basis can be constructed by
acting iteratively left mutations on the $R_a^{\ast}=\co (a)$.
Because the Calabi-Yau $M$ is realized as a hypersurface in the
$CP^{N-1}$ , we shall restrict the sets of bundles $\{S_a\}$, $\{R_a\}$ 
on the $M$.

Now we introduce a function $Z(S_m)$ associated with each bundle $S_m$
on $M$ as
\ba
&&Z(S_m):=(-1)^d\cdot 
\int_M \ch (S_m^{\ast})\sqrt{\hat{A}(TM)}\cdot e^{-t [D]}
\times 
\sqrt{\hat{K}(-[D])}
\times \exp \left(\sum_{m\geq 2}(-[D])^m \cdot
x_m\right)\non
&&\qquad\qquad \qquad (m=0,1,2,\cdots ,N-1)
\,.\nom
\ea
It can be expressed as an integral formula on $M$
\ba
&&Z(S_m)
=(-1)^{d+1}\cdot N^{-1} \cdot \sum^{N-1}_{k=1}
\alpha^{km}(\alpha^k-1)^N\cdot 
\int_M \frac{\alpha^k}{e^{[D]}-\alpha^k}
\cdot \sqrt{\hat{A}(TM)}\cdot e^{-t[D]}\non
&&\qquad\qquad \times 
\sqrt{\hat{K}(-[D])}
\times \exp \left(\sum_{m\geq 2}(-[D])^m \cdot
x_m\right)\,,\label{sheaf3}
\,
\,\,\,\,\,\,(m=0,1,2, \cdots ,N-1)\,.
\ea
We analyze a central charge of Calabi-Yau $d$-fold embedded 
in ${CP}^{N-1}$ in the analysis based on the  mirror symmetry.
The central charge $Z(\{{Q}^G\})$ in Eq.(\ref{sheaf2}) 
is labelled by a set of 
charges $\{{Q}^G_j\}$ at the Gepner point. 
They are transformed cyclically under 
the ${\bf Z}_N$ action 
around $\psi =0$, ${Q}^G_j\rightarrow {Q}^G_{j+1}$
$({Q}^G_{j+N}\equiv {Q}^G_j)$.
They also satisfy a relation 
$\dis \sum^{N-1}_{j=0} {Q}^G_j=0$ and we can decompose each 
$Q^G_j$ as 
${Q}^G_j=:{q}^G_j-{q}^G_{j-1}$. Then the 
central charge $Z$ is reexpressed as 
\ba
&&Z=N^{-1}\cdot (-1)\cdot \sum^{N-1}_{k=1}(\alpha^k-1)^N\cdot
\left(\sum^{N-1}_{j=0}{q}^G_j\alpha^{jk}\right)\non
&&\qquad \times
\int_M
\frac{\alpha^k}{e^{[D]}-\alpha^k}\cdot e^{-t[D]} 
\cdot \sqrt{\hat{A}(TM)}
\times 
\sqrt{\hat{K}(-[D])}
\times \exp \left(\sum_{m\geq 2}(-[D])^m \cdot
x_m\right)\,.\nom
\ea
By using the formula Eq.(\ref{sheaf3}) for $Z(S_m)$, we obtain a formula
of the $Z$ in terms of geometrical data
\ba
&&Z(\{q^G\})=(-1)^N\cdot \sum^{N-1}_{m=0}{q}^G_m 
\cdot Z(S_m)\non
&&\qquad =\sum^{N-1}_{m=0}{q}^G_m \int_M
\ch (S^{\ast}_m)
\cdot e^{-t[D]} \cdot \sqrt{\hat{A}(TM)} \non
&&\qquad 
\times 
\sqrt{\hat{K}(-[D])}
\times \exp \left(\sum_{m\geq 2}(-[D])^m \cdot
x_m\right)
\,.\label{eq2}
\ea
It is a linear combination of 
the $Z(S_m)$'s.
Each $Z(S_m)$ has information about 
a K-theory element of sheaves $S_m$ that is 
constructed through a restriction on the $M$. 
In the large radius region, 
fractional charges $\{Q_n\}$ in the fractional basis 
are defined as coefficients in 
an expansion of the above formula
\ba
&&Z(\{q^G\})=\sum_{n=0}^d\frac{t^{d-n}}{(d-n)!}\cdot (-1)^n
\cdot Q_{2(d-n)}\,,\non
&&Q_{2(d-n)}=
(-1)^{d}\cdot \sum^{N-1}_{m=0}{q}^G_m\int_M
\ch (S^{\ast}_m)\sqrt{\hat{A}(TM)}\cdot [D]^{d-n}\non
&&\qquad = (-1)^{d}\cdot \sum^{N-1}_{m=0}{q}^G_m
\int_{\gamma_{2n}}\ch (S^{\ast}_m)\sqrt{\hat{A}(TM)}\,,\nom
\ea
where we introduce a cycle ${\gamma_{2n}}$
as a dual of the $(2d-2n)$ form $[D]^{d-n}$.
We shall take an example with
${q}^G_m=\delta_{m,\ell}$, that is, 
\ba
Z=
\int_M \ch (S^{\ast}_{\ell})\sqrt{\hat{A}(TM)}\cdot e^{-t[D]}
\times 
\sqrt{\hat{K}(-[D])}
\times \exp \left(\sum_{m\geq 2}(-[D])^m \cdot
x_m\right)
\,.\nom
\ea
It is interesting that a pure D$(2d)$-brane is always 
labelled by a set of charges ${q}^G_m=\delta_{m,0}$, 
equivalently, ${Q}^G_0=+1$, ${Q}^G_1=-1$, 
${Q}^G_j=0$ $(j\neq 0,1)$. 

Next we compare our result in the sigma model with boundary states associated
with the Gepner model.
The boundary states are labelled by a set of integers 
$L_j$ $(j=1,2,\cdots ,N)$, $M$ and $S$ as 
$|\{L\};M;S\rangle$.
The integer $M$ is transformed under the ${\bf Z}_N$ monodromy
transformation  at the 
Gepner point as $M\rightarrow M+2$. 
On the other hand, 
the charge $q^G_m$ changes in such a transformation as
$q^G_m \rightarrow q^G_{m+1}$.
So the integral number $M$ turns out to be related with the charge
$q^G_m$.
For a trivial state $|\{L=0\};M=0;S=0\rangle $, we 
put an ansatz that an associated bundle is trivial 
one $\co$. It leads to a condition $q^G_m=\delta_{m,0}$. Then by
performing a ${\bf Z}_N$ transformation, we can 
construct a central charge $Z$
for a boundary state $|\{L=0\};M=2\ell;S=0\rangle $ 
\ba
&&Z(\{{q}^{G,0}_m\})=
\sum^{N-1}_{m=0}{q}^{G,0}_m\int_M \ch
      (S^{\ast}_m)\sqrt{\hat{A}(TM)}\cdot 
e^{-t[D]}\non
&&\qquad\qquad
\times 
\sqrt{\hat{K}(-[D])}
\times \exp \left(\sum_{m\geq 2}(-[D])^m \cdot
x_m\right)
\,,\non
&&{q}^{G,0}_m=\delta_{m,\ell}\,.\nom
\ea
For a boundary state $|\{L\};M=2\ell;S\rangle $ $(S=0,2)$, 
we obtain an associated central charge in the sigma model 
\ba
&&Z(\{{q}^{G}_m\})=
\sum^{N-1}_{m=0}{q}^{G}_m\int_M \ch
      (S^{\ast}_m)\sqrt{\hat{A}(TM)}\cdot 
e^{-t[D]}\non
&&\qquad\qquad 
\times 
\sqrt{\hat{K}(-[D])}
\times \exp \left(\sum_{m\geq 2}(-[D])^m \cdot
x_m\right)
\,,\non
&&{q}^G_m=
\frac{1}{N}\cdot (-1)^{\frac{S}{2}}\cdot 
\sum^{N-1}_{n=0}{q}^{G,0}_n
\sum^{N-1}_{k=1}\alpha^{k(n-m)}\times 
\prod^N_{j=1}\frac{\dis \sin \frac{\pi k (L_j +1)}{N}}{\dis \sin
      \frac{\pi k}{N}}\,,\non
&&{q}^{G,0}_n=\delta_{n,\ell}\,.\nom
\ea
Here we used the formula Eq.(\ref{zn}) of the charges when we switched on the 
$\{L_j\}$.
We can construct the central charge associated with the boundary state
of the Gepner model. 

Now let us return to the formula Eq.(\ref{sheaf2}) and 
study quantum effects in the $Z$.
They are contained (non-)perturbatively  
in the following terms
\ba
\sqrt{\hat{A}([D])}\cdot R([D])\cdot 
\sqrt{\hat{K}(-[D])}\cdot
e^{-t[D]}\times \exp \left(\sum_{n\geq 2}(-[D])^n \cdot
x_n\right)\,.\label{effect}
\ea
The term $\sqrt{\hat{A}}$ describes topological features of the 
associated bundle over the curved space. On the other hand, the 
$\hat{K}$ is expected to have its origin in the perturbative 
quantum corrections.
The $\hat{A}$ 
and the $\hat{K}$ are combined into a function
\ba
\sqrt{\hat{A}(+\rho)}\cdot \sqrt{\hat{K}(-\rho)}=
\frac{\dis \left[\Gam\!\left(1+\frac{\rho}{2\pi i}\right)\right]^N}
{\dis \Gam\!\left(1+\frac{N\rho}{2\pi i}\right)}
\,.\nom
\ea
Also this relation can be generalized to 
other Calabi-Yau cases, for an example, a Calabi-Yau $d$-fold realized
as complete intersections $M$ 
of $\ell$ hypersurfaces $\{p_{j}=0\}$ in
products of $k$ projective spaces ${\cal M}$
\ba
M:=\left(\matrix{{\bf P}^{n_1}(w_{1}^{(1)},\cdots ,w_{n_1+1}^{(1)})\cr
\vdots \cr
{\bf P}^{n_k}(w_{1}^{(k)},\cdots ,w_{n_k+1}^{(k)})
}
\right|\!\!
\left|
\matrix{d_{1}^{(1)}\cdots d_{\ell}^{(1)}\cr
\vdots \cr
d_{1}^{(k)}\cdots d_{\ell}^{(k)}}
\right)\,.\nom
\ea
The $d^{(i)}_j$ are degrees of the coordinates of 
${\bf P}^{n_i}(w_{1}^{(i)},\cdots ,w_{n_i+1}^{(i)})$ in the 
$j$-th polynomial $p_j$ ($i=1,2,\cdots ,k\,;\,j=1,2,\cdots ,\ell$).
When we introduce a function $a(v)$ with variables $v_i$ $(i=1,2,\cdots ,k)$
\ba
&&a(v):=
\frac{\dis \prod^{\ell}_{j=1}
\Gamma\!\left(1+\sum_{i=1}^k d_j^{(i)}v_i\right)}
{\dis \prod_{i=1}^k\prod_{j'=1}^{n_i+1}
\Gamma\!\left(1+w_{j'}^{(i)}v_i\right)}\,,\nom
\ea
a generating function of an $\hat{A}$ of the M is represented as
\ba
&&\hat{A}(\lambda):=
\prod_{i=1}^k\prod^{n_i+1}_{j'=1}\left(
\frac{\dis \frac{\lambda w^{(i)}_{j'}\rho_i}{2}}
{\dis \sinh \frac{\lambda w^{(i)}_{j'}\rho_i}{2}}\right)
\times
\prod^{\ell}_{j=1}\left(
\frac{\dis \sinh \frac{\lambda (d_{j}\rho)}{2}}
{\dis \frac{\lambda (d_{j}\rho)}{2}}\right)\non
&&\qquad =\sum_{m=1}\frac{(-1)^mB_m\cdot X_{2m}}{(2m)!(2m)}\lambda^{2m}\,,\non
&&d_j\cdot \rho :=
\sum_{i=1}^kd_j^{(i)}\rho_i\,,\non
&&X_n:=\sum_{i=1}^{k}\sum_{j'=1}^{n_i +1}(w^{(i)}_{j'}\rho_i)^n-
\sum_{j=1}^{\ell}\left(\sum_{i=1}^{k}d_j^{(i)}\rho_i\right)^n
\,,\non
&&d=-\ell +\sum^{k}_{i=1}n_i\,,\,\,\,\,(\mbox{dimension})\,.\nom
\ea
On the other hand, an associated $\hat{K}$ is defined by using the 
$X_n$s as
\ba
&&\hat{K}(\lambda):=
\exp\left[
+2\sum_{m=1}\frac{\zeta (2m+1)}{2m+1}\cdot \left(\frac{\lambda}{2\pi i}
\right)^{2m+1}\cdot X_{2m+1}\right]\,.\nom
\ea
The $\hat{A}(\lambda)$ and the $\hat{K}(\lambda)$ satisfy a relation
\ba
\sqrt{\hat{A}(+\lambda)}\cdot \sqrt{\hat{K}(-\lambda)}=
\frac{\dis \prod_{i=1}^k\prod_{j'=1}^{n_i+1}
\Gamma\!\left(1+\frac{\lambda w_{j'}^{(i)}\rho_i}{2\pi i}\right)}
{\dis \prod^{\ell}_{j=1}
\Gamma\!\left(1+\sum_{i=1}^k \frac{\lambda d_j^{(i)}\rho_i}{2\pi i}\right)}
\,.\nom
\ea
In our previous paper\cite{KS8}, 
we propose a conjecture that the 
$\sqrt{\hat{K}}$ could be interpreted as loop corrections of the 
sigma model. 
(For the quintic case, this statement is confirmed.)
If it is true for generic cases, this term might contain 
effects of perturbative corrections.
It seems interesting that these two terms 
are combined into a combination of Euler's gamma functions.

The remaining very important term is the 
$\dis \exp \left(\sum_{n\geq 2}(-[D])^n \cdot
x_n\right)$ in Eq.(\ref{effect}).
In order to study this function, we consider a set of functions 
$\{\check{\omega}_{\ell}\}$ defined in an expansion
\ba
&&e^{tv}\exp\left(\sum_{n\geq 2}v^nx_n\right)=
\sum_{\ell =0}v^{\ell}\check{\omega}_{\ell}\,.\nom
\ea
We collect these functions $\check{\omega}_{\ell}$ 
$(\ell =0,1,\cdots ,N-2)$ into a vector $\tilde{V}_0$
\ba
\tilde{V}_0=
\left(\matrix{
\check{\omega}_{0}&\check{\omega}_{1}& \cdots
&\check{\omega}_{N-2}}\right)\,.\nom
\ea
Here the $\check{\omega}_{\ell}$ can be interpreted as a paring of 
a B-cycle $\tilde{\gamma}_{\ell}$ and an A-model operator 
$\co^{(0)}$ associated with a $0$ form on the $M$
\ba
\check{\omega}_{\ell}=
\langle \tilde{\gamma}_{\ell}|\co^{(0)}\rangle\,.\nom
\ea
In our Calabi-Yau case, there are $(d+1)$ independent A-model operators 
$\co^{(m)}$ $(m=0,1,\cdots ,d)$ and we can define an associated 
vector $\tilde{V}_m$ for each $\co^{(m)}$
\ba
\tilde{V}_m=\left(
\matrix{
\langle \tilde{\gamma}_0 |\co^{(m)}\rangle &
\langle \tilde{\gamma}_1 |\co^{(m)} \rangle &
\cdots & 
\langle \tilde{\gamma}_d |\co^{(m)} \rangle }\right)\,,
\qquad (m=0,1,\cdots ,d)\,.\nom
\ea
Then we can introduce a matrix $\tilde{\Pi}$ 
by collecting these $(d+1)$ vectors
\ba
&&\tilde{\Pi}=\left(\matrix{
\tilde{V}_0\cr \tilde{V}_1\cr \vdots \cr \tilde{V}_d}\right)\,,\,\,\,
\tilde{\Pi}_{m,\ell}=\langle{\tilde{\gamma}_{\ell}| \co^{(m)}\rangle }\,.\nom
\ea
This matrix $\tilde{\Pi}$ satisfies a first order differential equation
\ba
&&\del_t\tilde{\Pi}=K\cdot \tilde{\Pi}\,,\label{connect}\\
&&K:=\left(\matrix{0 & \Kappa_0 & & & \cr 
            & 0        & \Kappa_1 & & \cr
            & & \ddots & \ddots & \cr
            & &  & 0 & \Kappa_{d-1} \cr
            & &  &  & 0} \right)\,.\nom
\ea
The $\Kappa_{\ell}$s are fusion couplings of A-model operators
defined as
\ba
&&\co^{(1)}\co^{(\ell -1)} = \Kappa_{j-1} \co^{(\ell )} \,\,\,\,\,\,\,\,
(1\leq \ell \leq d)\,\,\,,\non
&&\co^{(1)}\co^{(d)} = 0\,\,\,,\label{dfusion}\\
&&\Kappa_{m}=  {\dis {\partial_t} \frac{1}{\Kappa_{m-1}}{\partial_t} 
\frac{1}{\Kappa_{m-2}}{\partial_t} 
\cdots
{\partial_t} \frac{1}{\Kappa_{1}}
{\partial_t} \frac{1}{\Kappa_{0}}{\partial_t} \check{\omega}_{m+1}
}
\,\,\,\,\,\,\,\,(1\leq m \leq d-1)\,\,\,\,,\non
&& \Kappa_0 =1 \,\,\,.
\ea
When we put an initial data at a specific moduli point $t=t_i$, the 
$\tilde{\Pi}$ at a point $t=t_f$ is given by integrating the Eq.(\ref{connect})
formally
\ba
&&\tilde{\Pi}(t_f)
=\mbox{Pexp}\left(\int^{t_f}_{t_i}ds\,K(s)\right)\tilde{\Pi}(t_i)
\,.\nom
\ea
That is to say, the matrix $K$ plays a role of a connection on the 
moduli space. In other words, it 
induces a parallel transformation on the $t$-space.
When we impose a boundary condition 
$\tilde{\Pi}(t) \sim e^{t\cdot {\sf R}}$
on the $\tilde{\Pi}$ at $t=t_{\infty}$ 
$(\mbox{Im}t_{\infty}=+\infty)$, we obtain the $\tilde{\Pi}$  at a
generic point $t$
\ba
&&\tilde{\Pi}(t)=e^{t\cdot {\sf R}}\cdot \mbox{Pexp}
\left(\int^t_{t_{\infty}}ds\,
e^{-s\cdot {\sf R}}\tilde{K}(s)e^{+s\cdot {\sf R}}
\right)I\,,\non
&&{\sf R}:=\left(\matrix{0 & 1 & & & \cr 
            & 0        & 1 & & \cr
            & & \ddots & \ddots & \cr
            & &  & 0 & 1 \cr
            & &  &  & 0} \right)\,,\non
&&\tilde{K}:=K-{\sf R}\,.\nom
\ea
Thus the 
$\dis \exp \left(\sum_{n\geq 2}(-[D])^n \cdot
x_n\right)$ in Eq.(\ref{effect}) is 
evaluated by substituting $v=-[D]$ in the following formula
\ba
&&e^{tv}\exp\left(
{\sum_{n\geq 2}v^nx_n}\right)=\sum_{\ell =0}v^{\ell}\check{\omega}_{\ell}\non
&&=\left(\matrix{1 & t & \dis \frac{t^2}{2} & \cdots & \dis 
\frac{t^d}{d!}}\right)
\mbox{Pexp}
\left(\int^t_{t_{\infty}}ds\,
e^{-s\cdot R}\tilde{K}(s)e^{+s\cdot R}
\right)
\left(\matrix{1 \cr v \cr v^2 \cr \vdots \cr v^d}\right)\,\,.\nom
\ea
As a conclusion, we can interpret the 
$\dis \exp \left(\sum_{n\geq 2}(-[D])^n \cdot
x_n\right)$ in Eq.(\ref{effect}) as an effect of a parallel transport
through a path from the large radius point $t_{\infty}$ to a finite $t$ 
in the moduli space. Then the connection in the moduli space is
the $K$ that are defined by fusion couplings of A-model operators.

\section{Conclusions and Discussions}
In this article, we develop a method to construct the central charge in
the topological sigma model in the open string channel. 
First we analyzed quintic by using the associated prepotential $F$
and studied its monodromy properties.
For this quintic case, one can easily
construct associated canonical bases of period integrals because  
there is the prepotential for the model.
But one cannot expect to find analogs of prepotentials for other 
$d$-fold cases $(d>3)$. 
But we find that the essential part we learned in the analysis 
seems to be a factorizable property of the matrix ${\cal I}$ by 
a (triangular) matrix $S$ and a matrix $\Sigma$.
Under this consideration, we investigate the $N=7$ ($d=5$) case
concretely in section 4.

The basis $\Omega$ we pick here is a kind of symplectic one with 
an intersection matrix $\Sigma$. In order to obtain a set of canonical
basis $\Pi$, we determine the matrix $S$ which connects the $\Pi$ and
$\hat{\Pi}$. It allows us to construct the matrix ${\cal N}$ 
that transforms the $\Omega$ to the $\Pi$. Some topological invariants 
appear in the entries of the ${\cal N}$. Some parts of them are 
characterized by an A-roof genus of the $M$. But we cannot give 
geometrical interpretations to other entries in ${\cal N}$ explicitly.

Together with the data of the Gepner model, we calculate the 
charge vectors of the D-branes and the central charge. 
When a set of numbers $\{L_j\}$ is specified, a boundary state is 
constructed in the Gepner model. By calculating an 
associated charge $Q_G$ in the 
Gepner basis, we construct a formula of the $Z$ labelled by the set 
$\{L_j\}$. It is related to the boundary state $|\{L\};M;S\rangle$. 

In section 6, we investigate cycles associated with the sets of
periods. At the orbifold point $\psi =0$, they coincide with the 
susy cycles analyzed by Becker et. al.\cite{becker,hori}.
There we also analyze  intersection forms among them. 
In section 7, we reexpress our result of the $Z$ applicable in the 
large volume region of the $M$. We find that the $Z$ contains terms 
analogous to the Mukai vectors. 
They are interpreted as a product of Chern characters of 
bundles (sheaves) $S_m$s and a square root of the Todd class of the $M$. 
The set of $S_m$s is constructed as 
a dual basis of tautological line bundles of the 
ambient space $CP^{N-1}$ by a restriction on $M$.
In addition, there appear terms that 
encode perturbative and non-perturbative quantum corrections. 
The non-perturbative part is induced by 
a parallel transport
through a path from the large radius point $t_{\infty}$ to a finite $t$ 
in the moduli space. 
It turns out that the fusion coupling matrix $K$ of A-model operators 
plays a role of 
a connection on the moduli space.


\section*{Acknowledgment}
This work was supported by the Grant-in-Aid for 
Scientific Research from the Ministry of Education, Science and
Culture 10740117.

\newpage
\appendix

\section{Examples of $\sqrt{\hat{A}(\rho)}$}
We write down several examples of the $\sqrt{\hat{A}(\rho)}$ for the 
$d$-fold $M$ concretely. 
A function $\hat{A}(\rho)$ is defined as
\ba
&&\hat{A}(\rho)=
\left(\frac{\dis \frac{\rho}{2}}{\dis \sinh
\frac{\rho}{2}}\right)^N
\cdot 
\left(\frac{\dis \sinh \frac{N\rho}{2}}{\dis
\frac{N\rho}{2}}\right)\non
&&\qquad =\exp\left(\sum_{m=1}(-1)^m\rho^{2m}\cdot 
\frac{B_m}{(2m)!}\cdot \frac{X_{2m}}{2m}\right)\,.\nom
\ea

The coefficients $X_{\ell}=N-N^{\ell}$ are represented as some
combinations of Chern classes $c_{\ell}$s of the $M$ with $c_1=0$
\ba
&&c(\rho)=
1+\sum_{\ell \geq 1}\rho^{\ell}\frac{c_{\ell}}{N}
=\exp \left(\sum_{\ell \geq 1}(-1)^{\ell -1}\rho^{\ell}\cdot 
\frac{X_{\ell}}{\ell}\right)\,,\non
&&X_1=0\,,\,\,\,
X_2={\frac{-2\,c_{2}}{N}}\,,\,\,\,
X_3={\frac{3\,c_{3}}{N}}\,,\non
&&X_4={\frac{2\,{{c_{2}}^2}}{{N^2}}} - {\frac{4\,c_{4}}{N}}\,,\,\,\,
X_5={\frac{-5\,c_{2}\,c_{3}}{{N^2}}} + {\frac{5\,c_{5}}{N}}\,,\non
&&X_6={\frac{-2\,{{c_{2}}^3}}{{N^3}}} + {\frac{3\,{{c_{3}}^2}}{{N^2}}} + 
  {\frac{6\,c_{2}\,c_{4}}{{N^2}}} - {\frac{6\,c_{6}}{N}}\,,\non
&&X_7={\frac{7\,{{c_{2}}^2}\,c_{3}}{{N^3}}} - {\frac{7\,c_{3}\,c_{4}}{{N^2}}} - 
  {\frac{7\,c_{2}\,c_{5}}{{N^2}}} + {\frac{7\,c_{7}}{N}}\,,\non
&&X_8={\frac{2\,{{c_{2}}^4}}{{N^4}}} - {\frac{8\,c_{2}\,{{c_{3}}^2}}{{N^3}}} - 
  {\frac{8\,{{c_{2}}^2}\,c_{4}}{{N^3}}} + {\frac{4\,{{c_{4}}^2}}{{N^2}}}
  \non
&&\qquad + 
  {\frac{8\,c_{3}\,c_{5}}{{N^2}}} + {\frac{8\,c_{2}\,c_{6}}{{N^2}}} - 
  {\frac{8\,c_{8}}{N}}\,.\nom
\ea
In the central charge $Z$, there appears a function
$\sqrt{\hat{A}(\rho)}$. It is expanded in terms of the $\rho$ around $\rho=0$
\ba
&&\sqrt{\hat{A}(\rho)}=1+
\sum_{\ell =1}^d\beta_{\ell}\rho^{\ell}\,,\,\,\,
\beta_{2m+1}\equiv 0\,,\non
&&\beta_{2}=
{\frac{c_{2}}{24\,N}}\,,\,\,\,
\beta_{4}=
{\frac{7\,{{c_{2}}^2}}{5760\,{N^2}}} - {\frac{c_{4}}{1440\,N}}\,,\non
&&\beta_{6}=
{\frac{31\,{{c_{2}}^3}}{967680\,{N^3}}} - {\frac{{{c_{3}}^2}}{120960\,{N^2}}} - 
  {\frac{11\,c_{2}\,c_{4}}{241920\,{N^2}}} + {\frac{c_{6}}{60480\,N}}\,,\non
&&\beta_{8}=
{\frac{127\,{{c_{2}}^4}}{154828800\,{N^4}}} - 
  {\frac{11\,c_{2}\,{{c_{3}}^2}}{14515200\,{N^3}}} - 
  {\frac{113\,{{c_{2}}^2}\,c_{4}}{58060800\,{N^3}}} \non
&&\qquad + 
  {\frac{13\,{{c_{4}}^2}}{29030400\,{N^2}}} + 
  {\frac{c_{3}\,c_{5}}{2419200\,{N^2}}} 
+ {\frac{c_{2}\,c_{6}}{907200\,{N^2}}} - {\frac{c_{8}}{2419200\,N}}\,.\nom
\ea
We will summarize several examples of $\sqrt{\hat{A}(\rho)}$ 
for lower dimensional cases 
\ba
&&\hat{A}^{1/2}(\rho\,;\,N=3)=1\,,\,\,
\hat{A}^{1/2}(\rho\,;\,N=4)=1 + {\frac{{{\rho}^2}}{4}} \,,\non
&&\hat{A}^{1/2}(\rho\,;\,N=5)=1 + {\frac{5\,{{\rho}^2}}{12}} \,,\,\,\,
\hat{A}^{1/2}(\rho\,;\,N=6)=1 + {\frac{5\,{{\rho}^2}}{8}} 
- {\frac{11\,{{\rho}^4}}{384}} \,,\non
&&\hat{A}^{1/2}(\rho\,;\,N=7)=1 + {\frac{7\,{{\rho}^2}}{8}} 
- {\frac{21\,{{\rho}^4}}{640}} \,,\non
&&\hat{A}^{1/2}(\rho\,;\,N=8)=1 + {\frac{7\,{{\rho}^2}}{6}} - {\frac{7\,{{\rho}^4}}{240}} + 
  {\frac{229\,{{\rho}^6}}{1440}} \,,\non
&&\hat{A}^{1/2}(\rho\,;\,N=9)=1 + {\frac{3\,{{\rho}^2}}{2}} 
- {\frac{{{\rho}^4}}{80}} + {\frac{3233\,{{\rho}^6}}{10080}} \,,\non
&&\hat{A}^{1/2}(\rho\,;\,N=10)=1 + {\frac{15\,{{\rho}^2}}{8}} + {\frac{3\,{{\rho}^4}}{128}} + 
  {\frac{38861\,{{\rho}^6}}{64512}} -
  {\frac{3542981\,{{\rho}^8}}{3440640}} \,.\nom
\ea

\newpage

\section{Quintic Case}
We summarize monodromy matrices for the quintic case in the 
appendix for bases $\vpi$, $\Pi$, $\hat{\Pi}$, $\hat{\hat{\Pi}}$ and $\Omega$
\ba
&&\Pi ={\cal N}\cdot\Omega\,,\,\,\,
\Pi =m\cdot\vpi\,,\,\,\,
\vpi =P\cdot\Omega\,,\,\,\,\non
&&\Pi =S\cdot \hat{\Pi}\,,\,\,\,
\hat{\Pi}=U\hat{\hat{\Pi}}\,,\,\,\,
U_{m,n}=\left(\matrix{m\cr n}\right)\cdot (-1)^{m-n}\,,\non
&&U=\left(
\matrix{ 1 & 0 & 0 & 0 \cr -1 & 1 & 0 & 0 \cr 1 & -2 & 1 & 0 \cr -1 & 3 & 
    -3 & 1 \cr  } \right)\,,\,\,\,
m=\left(
\matrix{ 1 & 0 & 0 & 0 \cr -{\frac{2}{5}} & {\frac{2}{5}} & {\frac{1}
    {5}} & -{\frac{1}{5}} \cr {\frac{21}{5}} & -{\frac{1}{5}} & -{\frac{3}
     {5}} & {\frac{8}{5}} \cr 1 & -1 & 0 & 0 \cr  } 
\right)\,,\non
&&P\cdot V=\left(
\matrix{ 1 & 0 & 0 & 0 \cr 1 & {\frac{25}{12}} & 0 & 5 \cr -{\frac{23}
     {12}} & -{\frac{15}{4}} & -5 & -15 \cr -{\frac{23}{12}} & -{\frac{55}
     {12}} & -5 & -5 \cr  } 
\right)\,,\,\,\,
P=\left(
\matrix{ 0 & 0 & 0 & 1 \cr -5 & 0 & 0 & 1 \cr 15 & -5 & 0 & 1 \cr 
5 & -5 & 5 & -4 \cr  } \right)\,,\non
&&{\cal N}=m\cdot P\cdot  V=\left(
\matrix{ 1 & 0 & 0 & 0 \cr 0 & 1 & 0 & 0 \cr {\frac{25}{12}} & -{\frac{11}
     {2}} & -5 & 0 \cr 0 & -{\frac{25}{12}} & 0 & -5 \cr  } 
\right)
\,,\,\,\,
S=m\cdot P=
\left(
\matrix{ 0 & 0 & 0 & 1 \cr 0 & 0 & -1 & 1 \cr 0 & -5 & 8 & 
    -3 \cr 5 & 0 & 0 & 0 \cr  }  \right)\,.\nom
\ea
We write down monodromy matrices associated with singular points 
$\psi =0,1,\infty$ for various bases:
\begin{enumerate}
\item monodromy matrices associated with the $\psi=0$
\ba
&&A_{\varpi}=\left( 
\matrix{ 0 & 1 & 0 & 0 \cr 0 & 0 & 1 & 0 \cr -1 & -1 & -1 & 
    -1 \cr 1 & 0 & 0 & 0 \cr  } 
\right)\,,\,\,\,
A_{\Pi}=\left( 
\matrix{ 1 & 0 & 0 & -1 \cr -1 & 1 & 0 & 1 \cr 3 & 5 & 1 & -3 \cr 5 & -8 & 
    -1 & -4 \cr  } 
\right)\,,\non
&&A_{\hat{\Pi}}=\left(
\matrix{ -4 & 1 & 0 & 0 \cr -10 & 1 & 1 & 0 \cr -10 & 0 & 1 & 1 \cr 
    -5 & 0 & 0 & 1 \cr  }\right)\,,\,\,\,
A_{\Omega}=\left( 
\matrix{ 1 & {\frac{25}{12}} & 0 & 5 \cr -1 & -{\frac{13}{12}} & 0 & 
    -5 \cr {\frac{1}{2}} & {\frac{131}{144}} & 1 & {\frac{55}{12}} \cr -{
      \frac{1}{6}} & -{\frac{103}{144}} & -1 & -{\frac{23}{12}} \cr  } 
\right)\,,\non
&&A_{\hat{\hat{\Pi}}}=
\left(
\matrix{ -5 & 1 & 0 & 0 \cr -15 & 0 & 1 & 0 \cr -35 & 0 & 0 & 1 \cr 
    -71 & 4 & -6 & 4 \cr  } \right)\,.\nom
\ea
\item monodromy matrices associated with the $\psi=1$
\ba
&&P_{\varpi}=\left( 
\matrix{ 2 & -1 & 0 & 0 \cr 1 & 0 & 0 & 0 \cr -4 & 4 & 1 & 0 \cr 
    -4 & 4 & 0 & 1 \cr  } 
\right)\,,\,\,\,
P_{\Pi}=\left( 
\matrix{
   1 & 0 & 0 & 1 \cr 0 & 1 & 0 & 0 \cr 0 & 0 & 1 & 0 \cr 0 & 0 & 0 & 1 \cr  } 
\right)\,,\non
&&P_{\hat{\Pi}}=\left(
\matrix{
   1 & 0 & 0 & 0 \cr 5 & 1 & 0 & 0 \cr 5 & 0 & 1 & 0 \cr 5 & 0 & 0 & 1 \cr  }
\right)\,,\,\,\,
P_{\Omega}=\left( 
\matrix{ 1 & -{\frac{25}{12}} & 0 & -5 \cr 0 & 1 & 0 & 0 \cr 0 & -{\frac{125}
     {144}} & 1 & -{\frac{25}{12}} \cr 0 & 0 & 0 & 1 \cr  } 
\right)\,,\non
&&P_{\hat{\hat{\Pi}}}=
\left(
\matrix{
   1 & 0 & 0 & 0 \cr 5 & 1 & 0 & 0 \cr 15 & 0 & 1 & 0 \cr 35 & 0 & 0 & 1 \cr
    } \right)\,.\nom
\ea
\item monodromy matrices associated with the $\psi=\infty$ ($t\rightarrow t+1$)
\ba
&&T_{\varpi}=\left( 
\matrix{ 1 & 0 & 0 & 0 \cr 2 & 0 & 0 & -1 \cr -4 & 1 & 0 & 4 \cr -5 & -1 & 
    -1 & 3 \cr  } 
\right)\,,\,\,\,
T_{\Pi}=\left( 
\matrix{ 1 & 0 & 0 & 0 \cr 1 & 1 & 0 & 0 \cr -8 & -5 & 1 & 0 \cr 
    -5 & 3 & 1 & 1 \cr  } 
\right)\,,\non
&&T_{\hat{\Pi}}
=\left( 
\matrix{ 1 & -1 & 1 & -1 \cr 0 & 1 & -1 & 1 \cr 0 & 0 & 1 & 
    -1 \cr 0 & 0 & 0 & 1 \cr  } \right)\,,\,\,\,
T_{\Omega}=\left( 
\matrix{ 1 & 0 & 0 & 0 \cr 1 & 1 & 0 & 0 \cr {\frac{1}
    {2}} & 1 & 1 & 0 \cr {\frac{1}{6}} & {\frac{1}{2}} & 1 & 1 \cr  } 
\right)\,,\non
&&T_{\hat{\hat{\Pi}}}=
\left(
\matrix{ 4 & -6 & 4 & 
    -1 \cr 1 & 0 & 0 & 0 \cr 0 & 1 & 0 & 0 \cr 0 & 0 & 1 & 0 \cr  } 
\right)\,.\nom
\ea

\end{enumerate}

\end{document}